\documentstyle[]{article} 
\title{Implementation of Endomorphisms of the CAR~Algebra} 
\author{Carsten Binnenhei\thanks{Supported by the Deutsche
    Forschungsgemeinschaft (Sfb 288 ``Differentialgeometrie und
    Quantenphysik'')}~\thanks{E--mail: binnenhe@physik.fu-berlin.de}
    \\[1ex]Institut f\"ur Theoretische Physik der Freien
    Universit\"at Berlin\\Arnimallee 14, 14195 Berlin, Germany}
\date{February 1995}
\input amssym.def

\newtheorem{Def}{{\bf Definition}}[section]
\newtheorem{Th}[Def]{{\bf Theorem}}
\newtheorem{Lem}[Def]{{\bf Lemma}}
\newtheorem{Prop}[Def]{{\bf Proposition}}
\newtheorem{Cor}[Def]{{\bf Corollary}}

\newenvironment{Proof}{{\it Proof.}}{\nopagebreak\hspace*{\fill} $\Box$\\}
\newenvironment{Proofspc}{{\it Proof.}}{\nopagebreak\hspace*{\fill}
  $\Box$\vspace*{1.5ex}\\}

\newcommand{\aNN}{annihilation }
\newcommand{\AS}{antisymmetric}
\newcommand{\aS}{antisymmetric }
\newcommand{\ASS}{associat}
\newcommand{\AUT}{automorphism}
\newcommand{\aUT}{automorphism }
\newcommand{\BH}{bilinear Hamiltonian}
\newcommand{\bH}{bilinear Hamiltonian }
\newcommand{\BOG}{Bogoliubov}
\newcommand{\bOG}{Bogoliubov }
\newcommand{\BP}{basis projection}
\newcommand{\bP}{basis projection }
\newcommand{\CAR}{CAR~algebra}
\newcommand{\cAR}{CAR~algebra }
\newcommand{\CE}{conditional expectation}
\newcommand{\cE}{conditional expectation }
\newcommand{\CMP}[1]{{\em Commun. Math. Phys.} {\bf#1}}

\newcommand{\cOC}{continuous curve }
\newcommand{\CON}{connect}

\newcommand{\cOR}{corresponding }
\newcommand{\DEC}{decomposition}
\newcommand{\dEC}{decomposition }
\newcommand{\DIM}{dimension}
\newcommand{\dIM}{dimension }
\newcommand{\END}{endomorphism}
\newcommand{\eND}{endomorphism }
\newcommand{\FR}{Fredholm}
\newcommand{\fR}{Fredholm }
\newcommand{\FS}{Fock state}

\newcommand{\HS}{Hilbert--Schmidt}
\newcommand{\hS}{Hilbert--Schmidt }
\newcommand{\HSP}{Hilbert space}
\newcommand{\hSP}{Hilbert space }
\newcommand{\iFF}{if and only if }
\newcommand{\IMP}{implement}
\newcommand{\iMP}{implement }
\newcommand{\INC}{inclusion}
\newcommand{\iNC}{inclusion }
\newcommand{\INV}{invariant}
\newcommand{\iNV}{invariant }
\newcommand{\IRR}{irreducible}
\newcommand{\iRR}{irreducible }
\newcommand{\ISM}{isometr}
\newcommand{\ISO}{isomorphism}
\newcommand{\iSO}{isomorphism }

\newcommand{\oNB}{orthonormal basis }
\newcommand{\OP}{operator}
\newcommand{\oP}{operator }
\newcommand{\QEQ}{quasi--equivalen}
\newcommand{\QF}{quadratic form}
\newcommand{\qF}{quadratic form }
\newcommand{\QS}{quasi--free state}
\newcommand{\qS}{quasi--free state }
\newcommand{\REP}{representation}
\newcommand{\rEP}{representation }
\newcommand{\RUI}{Ruijsenaars}
\newcommand{\rUI}{Ruijsenaars }
\newcommand{\SDC}{selfdual \CAR}
\newcommand{\sDC}{selfdual \cAR}
\newcommand{\SG}{semigroup}
\newcommand{\sG}{semigroup }
\newcommand{\TR}{transformation}
\newcommand{\tR}{transformation }
\newcommand{\UNI}{unitar}
\newcommand{\WAT}{Watatani}
\newcommand{\wAT}{Watatani }

\newcommand{\0}{\!\circ\!}                    %
\newcommand{\1}{{\bf1}}                       %
\newcommand{\2}{{\oplus}}                     %
\newcommand{\3}{\beta}
\newcommand{\4}[1]{{\overline{#1}}}           %
\newcommand{\5}{{\gamma}}                     %
\newcommand{\6}{{\sigma}}                     %
\newcommand{\7}{{\Gamma}}                     %
\newcommand{\8}{{\alpha}}                     %
\newcommand{\9}{{\lambda}}                    %
\newcommand{\<}[1]{{\langle{#1}\rangle}}      %

\newcommand{\AD}{\mbox{\rm Ad}}               %
\newcommand{\B}[1]{{{\frak B}({#1})}}         
\newcommand{\CC}{{\Bbb C}}                    
\newcommand{\CK}{{\cal C}(\KK,\7)}            %
\newcommand{\CKJ}[1]{{\cal C}(\KK_{#1},\7_{#1})}
\newcommand{\DD}{{\cal D}}                    %
\newcommand{\EH}{\WO{\exp(b(H)/2)}}           %
\newcommand{\EV}{\WO{\exp(b(\LV)/2)}}         %
\newcommand{\FA}[1]{{\FF_a({#1})}}            %
\newcommand{\FF}{{\cal F}}                    %
\newcommand{\HH}{{\cal H}}                    %
\newcommand{\IK}[1]{{{\cal I}_{#1}({\KK},\7)}}%
\newcommand{\INK}[2]{{{\cal I}^{{#1}}_{#2}({\KK},\7)}} %
\newcommand{\IDX}{\mbox{\rm index\,}}           

\newcommand{\IFFF}{\Leftrightarrow}           
\newcommand{\IND}{\mbox{\rm\,ind\,}}           
\newcommand{\iND}{\mbox{\rm\scriptsize ind\,}}
\newcommand{\IN}{{I_{N_V}}}
\newcommand{\JK}[1]{{{\frak J}_#1({\KK})}}    %
\newcommand{\KK}{{\cal K}}                    %
\newcommand{\LL}[1]{{\Lambda_{{#1}}}}         %
\newcommand{\LU}{{\Lambda(U)}}                %
\newcommand{\LV}{{\Lambda(V)}}                %
\newcommand{\LW}{{\Lambda(W)}}                %
\newcommand{\MAT}[4]{%
  \left(\begin{array}{cc}{#1} & {#2} \\ {#3} & {#4} \end{array}\right)}
\newcommand{\mAT}[4]{{{#1}\,{#2} \choose {#3}\,{#4}}}
\newcommand{\MM}{{\cal M}}                  
\newcommand{\NN}{{\Bbb N}}                  
\newcommand{\OM}{\omega}
\newcommand{\PEME}{{\Psi_1(-\1)}}             %
\newcommand{\PHI}{{\varphi}}                  %
\newcommand{\PLV}{{{\cal P}_{L_V}}}           %
\newcommand{\PNV}{{\Psi_0(V)}}
\newcommand{\PME}{{\Psi(-\1)}}                %
\newcommand{\PS}{{{\frak P}_{L_V}}}           %
\newcommand{\PV}{{\hat\Psi(V)}}               %
\newcommand{\QK}{{\cal Q}(\KK,\7)}            %
\newcommand{\R}[1]{{\varrho_{#1}}}            %
\newcommand{\RAN}{{\rm ran\,}}
\newcommand{\rAN}{{\rm\scriptsize ran\,}}
\newcommand{\RE}{\mbox{\rm Re\,}}             %
\newcommand{\RR}{{\Bbb R}}                    %
\newcommand{\SIGS}{\mbox{\rm sign}\,\6}       %
\newcommand{\SSS}{{\cal S}}                  %
\newcommand{\TRA}{\mbox{\rm tr\,}}            %
\newcommand{\WO}[1]{{\::\!{#1}\!:\:}}         %
\newcommand{\WT}[1]{{\widetilde{#1}}}         %
\newcommand{\ZZ}{{\Bbb Z}}                    

\begin{document}
\maketitle
\begin{abstract}
  The \IMP ation of non--surjective \bOG\TR s in Fock states over \CAR
  s is investigated. Such a \tR is \IMP able by a \hSP of \ISM ies
  \iFF the well--known Shale--Stinespring condition is met. In this
  case, the \dIM of the \IMP ing \hSP equals the square root of the
  \wAT index of the \ASS ed \iNC of \CAR s, and both are determined by
  the \fR index of the \cOR one--particle \OP\@. Explicit expressions
  for the \IMP ing \OP s are obtained, and the \CON ed components of
  the \sG of \IMP able \TR s are described.
\end{abstract}
\section{Introduction}
\label{sec:INTRO}
The \IMP ation of \bOG {\em auto\/}morphisms of the algebra of
canonical anticommutation relations (CAR) by \UNI y \OP s on Fock
space is well--understood. Shale and Stinespring~\cite{SS65} have
proven that such an \aUT is \IMP able in a Fock \rEP \iFF the \cOR
one--particle \bOG\oP satisfies a certain \hS condition, and several
authors (e.g.~Friedrichs~\cite{Fri}, Berezin~\cite{Ber}, Labont\'e
\cite{Lab}, Fredenhagen~\cite{F77}, Klaus and Scharf~\cite{KS}, \rUI
\cite{R77a,R78}) have constructed the \IMP ing \UNI ies in terms of
\aNN and creation \OP s.

Here we tackle the problem of extending these results to the case of
\bOG{\em endo\/}morphisms. As suggested by the work of Doplicher and
Roberts~\cite{DR} on the theory of superselection sectors (see
\cite{H} for an overview), the appropriate generalization of `\IMP
ation of \AUT s by \UNI y \OP s' is `\IMP ation of \END s by \HSP s of
\ISM ies'. An \eND $\R{}$ is \IMP able in a \rEP$\pi$ of an arbitrary
C*--algebra \iFF $\pi\0\R{}$ is \UNI ily equivalent to a multiple of
$\pi$, and then the multiplicity equals the \dIM of an \IMP ing
\HSP\@. For \IRR~$\pi$, \IMP ability is tantamount to \QEQ ce of $\pi$
and $\pi\0\R{}$.

In the case of \bOG\END s and (\IRR) Fock \REP s of the \CAR, one may
apply the criterion for \QEQ ce of \QS s due to Powers and St\o rmer
\cite{PS} and Araki~\cite{A70} to conclude that a \bOG\eND is \IMP
able in the above sense \iFF the \cOR\bOG\oP fulfills the
Shale--Stinespring condition. The \dIM of the \IMP ing \hSP is then
given by the square root of the \wAT index~\cite{Wat} of the \ASS ed
\iNC of C*--algebras, and this index in turn equals $2^{-\iND V}$
where $-\IND V\in2\NN\cup\{\infty\}$ denotes the \fR index of the
\cOR\ISM ic \bOG\oP$V$. As shown by Longo~\cite{L}, an analogous
result holds in the theory of superselection sectors where the
statistical \dIM of a localized \eND coincides with the square root of
the Jones index of the \ASS ed \iNC of local algebras.

We derive explicit formulae for the \IMP ing \ISM ies (i.e.\ for
an \oNB of the \IMP ing \HSP) based on the work of \RUI~\cite{R78}.
For this purpose, we generalize the definition of \RUI' \oP $\LL{}$
(called the `\ASS e' of a given \bOG\OP) and obtain {\em one\/} \IMP
ing \ISM y $\Psi_0$ in terms of the Wick ordered exponential of the
unbounded \bH induced by $\LL{}$. A complete set of \IMP ing \ISM ies
may then be constructed by multiplying $\Psi_0$ with suitable partial
\ISM ies. In this way, the \IMP ing \hSP itself acquires a Fock space
structure, with $\Psi_0$ playing the role of the vacuum.

The set of \bOG\OP s $V$ fulfilling the Shale--Stinespring condition
(for a fixed Fock \REP) forms a topological \sG w.r.t.\ a suitably
chosen metric. By a result of Araki~\cite{A87}, the subgroup of \UNI
ies ($\IND V=0$) consists of two \CON ed components. We prove by
contrary that each subset of \bOG\OP s with fixed non--vanishing \fR
index is \CON ed.

Our interest in \IMP able \bOG\END s originates from the speculation
that they might serve to construct localized \END s for free Fermi
fields with non--abelian gauge groups~\cite{CB}. We intend to discuss
this idea in a subsequent paper. It should be mentioned that \bOG\TR s
have been successfully used in the construction of localized \END s in
conformal field theory models~\cite{MS,FGV,JMB1}.

This article is organized as follows. \CAR s, \bOG\TR s and \QS s are
introduced in Section~\ref{sec:PRE}. Throughout the paper Araki's
formalism of \SDC s~\cite{A68,A70,A87} is used which is equivalent to
the more familiar notion of complexified Clifford algebras over real
\HSP s~\cite{SS64}. However, Araki's approach has the advantage of
being complex--linear from the beginning. The usual description of a
\cAR by means of \aNN and creation \OP s enters through Fock \REP s of
the \SDC. In this section, we also compute \wAT indices of \INC s that
are induced by arbitrary \bOG\END s.

Implementability of \END s of C*--algebras is defined in
Section~\ref{sec:IMP}. We shortly discuss uniqueness of \IMP ing \OP s
and indices of \ASS ed \INC s. Then we turn to \CAR s and \bOG\END s.
We describe the \dEC of $\pi\0\R{}$ into cyclic sub\REP s where $\pi$
is a Fock \rEP and $\R{}$ a \bOG\END.
The already mentioned Powers--St\o rmer--Araki criterion then enables
us to prove the validity of the Shale--Stinespring condition in the
general case. We have included a new proof of a recent result of
B\"ockenhauer~\cite{JMB2} (\dEC of $\pi\0\R{}$ into \IRR s) in
Section~\ref{sec:DECREP} since we consider our proof to have some
interest on its own. 
We show that $\pi\0\R{}$ is equivalent to a multiple of either a
Fock \rEP or a direct sum of two inequivalent pseudo Fock \REP s,
depending on the index.

Section~\ref{sec:CON} contains the main result of our investigation,
namely the detailed construction of a complete set of \IMP ers for a
given \IMP able \END. In Section~\ref{sec:BH}, Wick ordered unbounded
\BH s and their Wick ordered exponentials are defined in a
\REP--dependent way with the help of unsmeared \aNN and creation \OP
s. Then commutation relations of these exponentials with \aNN and
creation \OP s are computed. The \ASS e $\LL{}$ is characterized by
intertwining properties of the corresponding exponential, but is not
unique. A complete set of \IMP ing \ISM ies is defined in
Section~\ref{sec:NORMFORM}. As a key to the proof of completeness, we
present a \dEC of $\R{}$ into a product of two simpler \TR s in
Section~\ref{sec:DEC}. This product \dEC also leads to an interesting
\dEC of \IMP ers.

Finally, we prove the aforementioned result on \CON edness in
Section~\ref{sec:STRUC}. Our argumentation parallels in part the
reasoning of Carey, Hurst and O'Brien in~\cite{CHOB} and relies on the
product \dEC developed in Section~\ref{sec:DEC}.
\section{Preliminaries}
\label{sec:PRE}
Let $\KK$ be an infinite--\DIM al complex \HSP\footnote{We are solely
  dealing with separable \HSP s in this article.} with a fixed
conjugation (i.e.\ anti\UNI y involution) $\7$, and let $\B{\KK}$ be
the algebra of bounded linear \OP s on $\KK$.  For $A\in\B{\KK}$ we set
$$\4{A}:=\7A\7.$$ 
Let ${\cal C}_0(\KK,\7)$ be the *--algebra, unique up to \ISO, which is
algebraically generated by the range of a linear embedding $B:\KK\to
{\cal C}_0(\KK,\7)$ with relations
\begin{equation} \label{SDCAR}
  \begin{array}{l}
    B(k)^*=B(\7 k),\\
    \{B(k)^*,B(k')\}=\<{k,k'}\1,\quad k,k'\in\KK.
  \end{array}
\end{equation}
Here $\{\,,\}$ denotes the anticommutator. ${\cal C}_0(\KK,\7)$ is
just the (complexified) Clifford algebra~\cite{SS64,Man} over the real
\hSP $\RE\KK:=\{k\in\KK\ |\ \7k=k\}$; conversely, given a real \HSP,
one may recover $\KK,\7$ (and $B$) by complexification (details are
in~\cite{CB}). There is a unique C*--norm on ${\cal C}_0(\KK,\7)$
(which fulfills $\|B(k)\|^2=\frac{1}{2}(\|k\|^2+(\|k\|^4- |\<{k,\7
  k}|^2)^{1/2})$), and completion in this norm yields a simple
C*--algebra $\CK$, namely Araki's {\em \SDC\/} over
$(\KK,\7)$~\cite{A68,A70,A87}.

{\em\bOG\TR s\/} are precisely the unital *--\END s of $\CK$ that
leave $\KK$ \INV\@. Put differently, every \ISM y $V\in\B{\KK}$ that
commutes with $\7$ (and therefore restricts to a real--linear \ISM y
of $\RE\KK$) induces a unital, \ISM ic *--\eND $\R{V}$ of $\CK$ through
$$\R{V}(B(k))=B(Vk),\quad k\in\KK.$$ Such \ISM ies are called
{\em\bOG\OP s}, and the \sG of \bOG\OP s is denoted by
$$\IK{}:=\{V\in\B{\KK}\ |\ V^*V=\1,\ \4{V}=V\}.$$ The map
$V\mapsto\R{V}$ is a unital \iSO from $\IK{}$ onto the \sG of \bOG\END
s; for fixed $A\in\CK$, the map $V\mapsto\R{V}(A)$ is continuous
w.r.t.\ strong topology on $\IK{}$ and norm topology on $\CK$.

Let $V\in\IK{}$. Since $\RAN V$ is closed and $\ker V=\{0\}$, $V$ and
$V^*$ are semi--\fR \OP s in the sense of Kato~\cite{K} and have
well--defined \fR indices. The map
$$\IK{}\to\NN\cup\{\infty\},\quad V\mapsto\IND V^*=-\IND V=\dim\ker
V^*$$ is a surjective homomorphism of \SG s ($0\!\in\!\NN$ by convention).
Hence $\IK{}$ is the disjoint union of subsets
\begin{equation} \label{INK}
   \IK{}=\bigcup_{n\in\NN\cup\{\infty\}}\INK{n}{},\qquad
   \INK{n}{}:=\{V\in\IK{}\ |\ \IND V^*=n\}. 
\end{equation}
Note that $\R{V}$ is an \aUT\iFF$V\in\ \INK{0}{}$, the group of \UNI y
\bOG\OP s, in which case we prefer to use the symbol ``$\8$'' instead
of ``$\R{}$''. For $V_1,V_2\in\INK{n}{}$ there exists $U\in\INK{0}{}$
with $V_1=UV_2$. Such $U$ has the form $U=V_1^{}V_2^*+u$ where $u$ is a
partial \ISM y with $(\ker u)^\bot=\ker V_2^*,\ \RAN u=\ker V_1^*$, and
$u=\4{u}$. We may express this in a more sophisticated way by saying
that $\INK{0}{}$ acts on $\IK{}$ by left multiplication, that the
orbits of this action are just the sets $\INK{n}{}$, and that the
stabilizer of $V\in\INK{n}{}$ is isomorphic to $O(n)$ (the orthogonal
group of an $n$--\DIM al real \HSP). 

Next we describe the set of states we are interested in. A state $\OM$
over $\CK$ is called {\em quasi--free\/}~\cite{A70} if its n--point
functions have the form
\begin{eqnarray*}
   \OM(B(k_1)\cdots B(k_{2m+1})) &=& 0, \\
   \OM(B(k_1)\cdots B(k_{2m})) &=& (-1)^\frac{m(m-1)}{2}\sum_{\6}
      \SIGS\ \OM\Big(B(k_{\6(1)})
      B(k_{\6(m+1)})\Big)\cdots \OM\Big(B(k_{\6(m)}) B(k_{\6(2m)})\Big)
\end{eqnarray*}
where the sum runs over all permutations $\6$ satisfying
$\6(1)<\ldots<\6(m)$ and $\6(j)<\6(j+m),\ j=1,\ldots,m$.
Therefore \QS s are completely determined by their two--point functions,
and we have a bijection between the convex set
$$\QK:=\{S\in\B{\KK}\ |\ 0\le S\le\1,\ \4{S}=\1-S\}$$
and the (non--convex) set of \QS s given by
$$S\mapsto\OM_S,\quad\OM_S(B(k)^*B(k'))=\<{k,Sk'}.$$
The following lemma is immediate.
\begin{Lem}\label{lem:OP}
   The \sG of \bOG\END s acts from the right on the set of \QS s by
   $\OM\mapsto\OM\0\R{}$, $\IK{}$ acts from the right on
   $\QK$ by $S\mapsto V^*SV$, and $$\OM_S\0\R{V}=\OM_{V^*SV}.$$
\end{Lem}

Projections in $\QK$ are called {\em \BP s\/} and the \cOR states
{\em\FS s\/}; the latter are precisely the {\em pure\/} \QS
s~\cite{MRT}. The group of \bOG\AUT s acts transitively on the set of
Fock states as $\INK{0}{}$ acts transitively on the set of \BP s.
Note that for a \bP $P$, the complementary (basis) projection is simply
given by $\4{P}$. Since $\OM_P(B(k)^*B(k))=0$ if $k\in\4{P}(\KK)$, the
elements of $B(\4{P}(\KK))$ (resp.\ $B(P(\KK))$) correspond to \aNN
(resp.\ creation) \OP s in the state $\OM_P$. A (faithful and \IRR)
GNS \rEP $\pi_P$ for $\OM_P$ is given by
$$\pi_P(B(k)):=a(Pk)^*+a(P\7k)$$
on the \aS Fock space $\FA{P(\KK)}$ over $P(\KK)$ with the
usual Fock vacuum $\Omega_P$ as cyclic vector and \aNN \OP s
$a(f),\ f\in P(\KK)$. In a Fock \rEP $\pi_P$, a \bOG\eND
$\R{V}$ induces the \tR 
\begin{equation} \label{TR}
  a(f)\mapsto a_V(f):= a(PVPf)+a(PV\4{P}\7f)^*,\quad f\in P(\KK),
\end{equation}
which shows the \CON ion to the (state--dependent) description of \bOG\TR
s by pairs of \OP s $(PVP,PV\4{P}\7)$ as preferred by some authors
(e.g.~\cite{BR2}).

Given a \bP $P$, a state over $\CK$ is said to be {\em gauge \iNV\/}
if it is \iNV under the one--parameter group of \bOG\AUT s
$(\8_{U_\9})_{\9\in\RR}$ with $U_\9:=e^{i\9}P+e^{-i\9}\4{P}
\in\INK{0}{}$. As follows from Lemma~\ref{lem:OP}, a \qS $\OM_S$ is
gauge \iNV \iFF $[P,S]=0$.

The so--called {\em central state\/} $\OM_{1/2}$~\cite{SS64,Man,A70} is
the unique tracial state over $\CK$. By uniqueness, $\OM_{1/2}$ is
\iNV under all unital *--\END s of $\CK$.

Now suppose we have an orthogonal \dEC $\KK=\KK_1\2\KK_2$ into
$\7$--\iNV closed subspaces with $\KK_2$ finite \DIM al\footnote{If
  $\dim\KK_2$ is odd, then $\CKJ{2}$ is not uniquely determined by
  (\ref{SDCAR}); in addition, one requires it to have non--trivial
  center (see~\cite{A87}).}. Set $\7_j:=\7|_{\KK_j}$ and regard
$\CKJ{j},\ j=1,2$ as subalgebras of $\CK$. Then $\CK$ is canonically
isomorphic to the $\ZZ_2$--graded tensor product of $\CKJ{1}$ and
$\CKJ{2}$~\cite{Hm} through identification of $A_1\otimes A_2$ with
$A_1\cdot A_2\in\CK,\ A_j\in\CKJ{j}$ (here $\ZZ_2=\{0,1\}$, and the
grading is induced by $\8_{-1}$, i.e.\ $\CK=\CK_0\2\CK_1,\ \CK_g:=\{A\ 
|\ \8_{-1}(A)=(-1)^gA\}$).
Hence all elements of $\CK$ are finite sums of elements $A_1A_2$ as
above, and we have a well--defined linear mapping
$$E:\CK\to\CKJ{1},\quad A_1A_2\mapsto A_1\OM_{1/2}(A_2),\quad A_j\in\CKJ{j}.$$
\begin{Lem}\label{lem:INDE}
   $E$ is a faithful \cE from $\CK$ onto $\CKJ{1}$ with
   \wAT index
   $$\IDX E=\dim\CKJ{2}=2^{\dim\KK_2}.$$
\end{Lem}
\begin{Proof}
   We first show $E(A^*)=E(A)^*,\ A\in\CK$. By linearity, it suffices
   to check this for elements of the form $A=A_1A_2$ with
   $A_j\in\CKJ{j}$ homogeneous. By use of the anticommutation relations, 
   \begin{eqnarray*}
      E(A^*) &=& (-1)^{\deg A_1\deg A_2}E(A_1^*A_2^*) \\
             &=& (-1)^{\deg A_1\deg A_2}A_1^*\OM_{1/2}(A_2^*) \\
             &=& A_1^*\4{\OM_{1/2}(A_2)}\qquad(\mbox{since }\deg
                 A_2\neq0\mbox{ implies }\OM_{1/2}(A_2^*)=0) \\
             &=& E(A)^*.
   \end{eqnarray*}
   Hence $E$ is positive. Now let $A,B_1,C\in\CKJ{1}$ and
   $B_2\in\CKJ{2}$ be given, with $B_2$ and $C$ homogeneous. Then
   \begin{eqnarray*}
      E(AB_1B_2C) &=& (-1)^{\deg B_2\deg C}E(AB_1CB_2) \\
                  &=& (-1)^{\deg B_2\deg C}AB_1C\OM_{1/2}(B_2) \\
                  &=& AE(B_1B_2)C.
   \end{eqnarray*}
   By linearity, $E(ABC)=AE(B)C$ for $A,C\in\CKJ{1},\ B\in\CK$, so $E$
   is a \CE.

   To compute the \wAT index~\cite{Wat} of $E$ we need a `quasi--basis',
   i.e.\ a finite subset $\{B_\3\}\subset\CK$ fulfilling
   \begin{equation} \label{QB}
      \sum_\3E(AB_\3)B_\3^*=A,\quad A\in\CK.     
   \end{equation}
   $\IDX E$ is then defined as $\IDX E:=\sum_\3B_\3B_\3^*$ and does
   not depend on the choice of quasi--basis. The existence of a
   quasi--basis also guarantees faithfulness of $E$.

   Here we may obtain a quasi--basis as follows. Let
   $\{b_1,\ldots,b_n\}$ be an \oNB for $\KK_2$ consisting of
   $\7$--\iNV vectors ($n<\infty$ by assumption). Let $I_n$ denote the
   set of $2^n$ multi--indices $\3=(\3_1,\ldots,\3_l)$ obeying
   \begin{equation} \label{I}
      0\leq l\leq n,\quad1\leq\3_1<\ldots<\3_l\leq n\quad
      (\3:=0\mbox{ for }l=0).
   \end{equation}
   Set $B_j:=\sqrt{2}B(b_j)$ for $j=1,\ldots,n$ and
   $B_\3:=B_{\3_1}\cdots B_{\3_l}$ for $\3\in I_n$ ($B_0:=\1$). 

   We claim that $(B_\3)_{\3\in I_n}$ is a quasi--basis for $E$ (by
   construction, it is a basis for $\CKJ{2}$). Note that
   $\{B_j,B_m\}=2\delta_{jm}\1,\ j,m=1,\ldots,n$, and
   $B_\3^*=(-1)^{l(l-1)/2}B_\3$ if $\3=(\3_1,\ldots,\3_l)\in I_n$.
   Furthermore $\OM_{1/2}(B_\3)=\delta_{\30}$~\cite{SS64,A70}, hence
   $\OM_{1/2}(B_\3^*B_\5)=\delta_{\3\5}$. Again by linearity, it
   suffices to consider elements of the form $A=A_1B^*_\3,\ 
   A_1\in\CKJ{1},\ \3\in I_n$. We have
   $$\sum_{\5\in I_n}E(AB_\5)B_\5^*=A_1\sum_{\5\in
      I_n}\OM_{1/2}(B_\3^*B_\5)B_\5^*=A_1B_\3^*=A.$$ 
   Therefore $(B_\3)_{\3\in I_n}$ is a quasi--basis for $E$, and using
   $B_\3B_\3^*=\1$ we get
   $$\IDX E=\sum_{\3\in I_n}B_\3B_\3^*=2^n\1,\quad n=\dim\KK_2.$$
\end{Proof}
Next we show that $E$ is the \cE with minimal index, so the index of
the \iNC of simple C*--algebras $\CKJ{1}\subset\CK$ equals
$$[\CK:\CKJ{1}]=\IDX E=2^{\dim\KK_1^\bot}.$$
\begin{Lem}\label{lem:MIN}
   $E$ is the unique minimal \cE from $\CK$ onto $\CKJ{1}$.
\end{Lem}
\begin{Proof}
   Following \WAT~\cite{Wat} we have to show 
   \begin{equation} \label{MIN}
      \IDX E\cdot E(A)=\sum_{\3\in I_n}B_\3AB_\3^*
   \end{equation}
   for $A\in\CKJ{1}^c$, the C*--algebra of elements of $\CK$ that
   commute with all elements of $\CKJ{1}$. We claim that $\CKJ{1}^c$
   equals $\CKJ{2}_0$, the even subalgebra of $\CKJ{2}$. Indeed,
   writing $A=\sum_\3A_\3B_\3$ with $A_\3\in\CKJ{1}$, all $A_\3$ have
   to commute with the elements of $\CKJ{1}_0$.
   Let $P$ be a \bP of $(\KK_1,\7_1)$ and $\PME$ a \UNI y \IMP ing
   $\8_{-1}$ in $\pi_P$ (which exists due to invariance of $\OM_P$
   under $\8_{-1}$ and is unique up to a phase) then
   $\pi_P(\CKJ{1}_0)''=\{\PME\}'$.  It follows that
   $\pi_P(A_\3)\in\pi_P(\CKJ{1}_0)'=\{\PME\}''=
   \mbox{span\,}\{\1,\PME\}$. But since $\8_{-1}$ is not inner
   \cite{SS65,A70,A87}, we have $\PME\not\in\pi_P(\CKJ{1})$. Thus
   $A_\3\in\CC\1$ and $A\in\CKJ{2}_0$.
   
   It suffices to prove (\ref{MIN}) for $A=B_\5,\
   \5=(\5_1,\ldots,\5_l)\in I_n,\ l$ even (the case $A=\1$ is clear by
   definition of $\IDX E$). In the following computation we use the
   notation $\3\cap\5:=\{\3_1,\ldots,\3_r\}\cap\{\5_1,\ldots,\5_l\}$
   if $\3=(\3_1,\ldots,\3_r)\in I_n$. $\3'\in I_n$ will then denote
   the multi--index whose entries are the elements of
   $\{\3_1,\ldots,\3_r\}\backslash(\3\cap\5)$.
   \begin{eqnarray*}
      \sum_{\3\in I_n}B_\3AB_\3^* &=&
         \sum_{m=0}^l\sum_{1\leq j_1<\ldots<j_m\leq l}
         \sum_{\3,\ \3\cap\5=\atop\{\5_{j_1},\ldots,\5_{j_m}\}}
         B_\3B_\5B_\3^* \\
      &=& \sum_{m=0}^l\sum_{j_1<\ldots<j_m}
         \sum_{\3\cap\5=\atop\{\5_{j_1},\ldots,\5_{j_m}\}}
         B_{\3'}B_{\5_{j_1}}\cdots B_{\5_{j_m}}B_\5(B_{\5_{j_1}}\cdots
         B_{\5_{j_m}})^*B_{\3'}^* \\ 
      &=& \sum_{m=0}^l\sum_{j_1<\ldots<j_m}
         \sum_{\3\cap\5=\atop\{\5_{j_1},\ldots,\5_{j_m}\}}
         (-1)^mB_\5\underbrace{B_{\3'}B_{\3'}^*}_{\1} 
         \underbrace{B_{\5_{j_1}}\cdots B_{\5_{j_m}}
         (B_{\5_{j_1}}\cdots B_{\5_{j_m}})^*}_{\1} \\
      &=& B_\5\sum_{m=0}^l(-1)^m{l\choose m}\cdot2^{n-l}\ =\
         2^{n-l}B_\5(-1+1)^l\ =\ 0.
   \end{eqnarray*}
   But we also have $E(B_\5)=\OM_{1/2}(B_\5)=0$ if $\5\neq0$.
\end{Proof}

Let us return to \bOG\TR s. The possible ranges of \bOG\OP s are just
the infinite--\DIM al $\7$--\iNV closed subspaces of $\KK$, and for
$V\in\IK{}$, we may identify $\R{V}(\CK)$ with ${\cal C}(\RAN V,\7|_{\RAN
 V})$. Thus we have just seen that 
$$[\CK:\R{V}(\CK)]=2^{\iND V^*}$$
if $\IND V^*<\infty$, and this causes us to assign to each \bOG\oP a number
\begin{equation} \label{DV}
  d_V:=2^{\frac{1}{2}\iND V^*}\leq\infty,\quad V\in\IK{}
\end{equation}
analogous to the statistical \dIM in the theory of superselection
sectors~\cite{L}. $d$ is obviously multiplicative
$$d_{VV'}=d_Vd_{V'}.$$
Note that $d_V$ is defined without reference to any \REP, but if
$\R{V}$ happens to be \IMP able in a Fock \REP, then $d_V$ shows up
as the \dIM of the \IMP ing \HSP. More generally, we shall see
in Section~\ref{sec:DECREP} that the \REP s $\pi_P\0\R{V}$ (with $P$ a
\bP and $V$ an arbitrary \bOG\OP) split into $d_V$ resp.\ 
$\sqrt{2}d_V$ \IRR s if $\IND V^*$ is even resp.\ odd
(cf.~\cite{JMB2}). Also note that the \CE s $E$ defined above allow the
definition of {\em left inverses\/}~\cite{H} $\R{}^{-1}\0E$ for
\bOG\END s. More explicitly, for a \bOG\eND $\R{V}$, a left
inverse $\Phi_V$ is given by
$\Phi_V(A_1A_2):=\R{V}^{-1}(A_1)\OM_{1/2}(A_2)$ if $A_j\in\CKJ{j},\ 
\KK_1:=\RAN V,\ \KK_2:=\ker V^*$.

An essential ingredient for our analysis in Section~\ref{sec:IMPDEC}
will be the criterion for \QEQ ce of \QS s as derived by Powers and
St\o rmer~\cite{PS} for gauge \iNV states and generalized by
Araki~\cite{A70}.  By definition, two states $\OM,\OM'$ are \QEQ t
(denoted by ``$\approx$'') if they induce \QEQ t GNS--\REP s. Now let
$\JK{p}$ be the trace ideal
$$\JK{p}:=\{A\in\B{\KK}\ |\ \|A\|_p<\infty\},\quad 1\le p<\infty$$
with trace norm $\|A\|_p:=(\TRA(|A|^p))^{1/p}$, and let
$S,S'\in\QK$. The statement is
\begin{equation} \label{QEQ}
  \OM_S\approx\OM_{S'}\iff S^{1/2}-S'^{1/2}\in\JK{2}.
\end{equation}
It has been observed by Powers~\cite{P87} that this criterion may be
simplified if one of the \OP s $S,S'$ is a projection. Namely, if $P$
is a \BP, then
\begin{equation} \label{PSC}
  \OM_P\approx\OM_S\iff \4{P}S\4{P}\in\JK{1}.
\end{equation}
\section{Implementability and Equivalence of Representations}
\label{sec:IMPDEC}
The famous result of Shale and Stinespring~\cite{SS65} asserts that a
\bOG\aUT $\8_V,\ V\in\INK{0}{}$, is \UNI ily \IMP able in a Fock \rEP
$\pi_P$ \iFF 
\begin{equation} \label{SS}
  [P,V]\in\JK{2}. 
\end{equation}
`Unitarily \IMP able' stands for the existence of a \UNI y \oP $\Psi$
on Fock space fulfilling $\AD\Psi\0\pi_P=\pi_P\0\8_V$ where
$$(\AD\Psi)(X):=\Psi X\Psi^*$$ (in the following, we shall use the
notation $\AD\Psi$ also for partially \ISM ic $\Psi$).  Note that the
Shale--Stinespring condition immediately follows from (\ref{QEQ}) (or
(\ref{PSC})). In fact, existence of $\Psi$ is equivalent to \QEQ
ce of the \iRR\REP s $\pi_P$ and $\pi_P\0\8_V$.  Since $\pi_P\0\8_V$
is a GNS--\rEP for $\OM_P\0\8_V=\OM_{V^*PV}$ (see Lemma~\ref{lem:OP}),
$\pi_P\approx\pi_P\0\8_V$ \iFF $P-V^*PV=V^*[V,P]\in\JK{2}$ by
(\ref{QEQ}) (remember that $P$ and $V^*PV$ are projections).
 
We shall show first that an \eND $\R{V}$ is \IMP able in a Fock \rEP
$\pi_P$ (in an appropriate sense) \iFF (\ref{SS}) holds. Later we
shall study the action of the group of \IMP able \AUT s on the \sG of
\END s (with finite index). This will lead us to a description of
equivalence classes of \REP s $\pi_P\0\R{V}$.
\subsection{Implementability of Endomorphisms}
\label{sec:IMP}
To generalize the notion of \IMP ability to the case of \END s we
adopt ideas of Doplicher and Roberts~\cite{DR}. The \UNI y \IMP er
$\Psi$ above gets thereby replaced by a set of \ISM ies fulfilling the
relations of a Cuntz algebra~\cite{C}. We give a definition for
arbitrary C*--algebras.
\begin{Def}\label{def:IMP}
   A *--\eND $\R{}$ of a C*--algebra $\frak A$ is {\em(\ISM ically) \IMP
   able\/} in a \rEP $(\pi,\HH)$ if there exists a (possibly finite)
   sequence $(\Psi_n)_{n\in I}$ in $\B{\HH}$ with relations
   \begin{equation} \label{CUNTZ}
      \Psi_m^*\Psi_n=\delta_{mn}\1,\qquad 
      \sum\limits_{n\in I}\Psi_n\Psi_n^*=\1\footnotemark,
   \end{equation}\footnotetext{w.r.t.\ strong topology if $I$ is 
      infinite}\addtocounter{footnote}{-1}which \IMP s $\R{}$ by
   \begin{equation} \label{IMPEND}
      \pi\0\R{}=\sum_{n\in I}\AD\Psi_n\0\pi\footnotemark.
   \end{equation}
\end{Def}
$\HH$ then decomposes into the orthogonal direct sum of the ranges of
the \ISM ies $\Psi_n$, and $\pi\0\R{}$ decomposes into sub\REP s
$\pi\0\R{}|_{\RAN\Psi_n}$, each of them \UNI ily equivalent to $\pi$. But the
converse is also true, i.e.\ $\R{}$ is \IMP able in $\pi$ \iFF
$\pi\0\R{}$ is equivalent to a multiple of $\pi$. For \iRR $\pi$ this reads
\begin{equation} \label{IMPQEQ}
  \R{}\mbox{ is \IMP able in }\pi\iff\pi\0\R{}\approx\pi.
\end{equation}

By (\ref{CUNTZ}), we may regard the \IMP ing \ISM ies $(\Psi_n)_{n\in
I}$ as an \oNB of the \hSP $H:=\4{\mbox{span\,}(\Psi_n)}$ in
$\B{\HH}$ with scalar product given by $\Psi^*\Psi'=\<{\Psi,\Psi'}\1$
(this scalar product induces the usual
\oP norm). Every element $\Psi$ of $H$ is 
an intertwiner from $\pi$ to $\pi\0\R{}$:
\begin{equation} \label{INTER}
  \Psi\pi(A)=\pi(\R{}(A))\Psi,\quad A\in{\frak A}.
\end{equation}
Note that $H$ coincides with the space of intertwiners from $\pi$ to
$\pi\0\R{}$ \iFF$\pi$ is \IRR\@. If $\pi$ is reducible, there may
exist several \HSP s \IMP ing $\R{}$, mutually related by \UNI ies in
$\pi(\R{}({\frak A}))'$. More precisely, if $(\Psi_n)_{n\in I}$ and
$(\Psi_n')_{n\in I}$ both \iMP$\R{}$ in $\pi$ (we may choose the same
index sets), then $\Psi:=\sum_n\Psi_n'\Psi_n^*$ is a \UNI y in
$\pi(\R{}({\frak A}))'$, and $\Psi_n'=\Psi\Psi_n$. Conversely, given
$(\Psi_n)_{n\in I}$ and a \UNI y $\Psi\in\pi(\R{}({\frak A}))'$,
$(\Psi\Psi_n)_{n\in I}$ is a set of \IMP ing \ISM ies
(cf.~\cite{F91}).

An \IMP able \eND $\R{}$ gives rise to normal *--\END s
$\R{H}:=\sum_{n\in I}\AD\Psi_n$ of $\B{\HH}$, and one finds~\cite{L} 
$$[\B{\HH}:\R{H}(\B{\HH})]=d_\R{}^2$$ where $d_\R{}:=\dim H$ does not
depend on the choice of $H=\4{\mbox{span\,}(\Psi_n)}$. Let us outline
the computation of the index in the setting of \wAT(cf.\ the proofs of
Lemmas~\ref{lem:INDE},~\ref{lem:MIN}) for the case $d_\R{}<\infty$.
$\Phi_H:=d_\R{}^{-1}\sum_n\AD\Psi_n^*$ is a left inverse for $\R{H}$
yielding the \cE$E_H:=\R{H}\0\Phi_H$ from $\B{\HH}$ onto
$\R{H}(\B{\HH})$.                            
$(\sqrt{d_\R{}}\Psi_n^*)_{n=1,\ldots,d_\R{}}$ is a quasi--basis (cf.\ 
(\ref{QB})) for $E_H$, hence $\IDX
E_H=d_\R{}\sum_n\Psi_n^*\Psi_n=d_\R{}^2$. To show minimality of $E_H$,
one must check (\ref{MIN}) $d_\R{}E_H(A)=\sum_l\Psi_l^*A\Psi_l$ for
$A\in\R{H}(\B{\HH})'$. But
$\R{H}(\B{\HH})'=\mbox{span\,}\{\Psi\Psi'^*\ |\ \Psi,\Psi'\in
H\}\cong\B{H}$, and $d_\R{}E_H(\Psi_m\Psi_n^*)=\delta_{mn}\1=
\sum_l\Psi_l^*(\Psi_m\Psi_n^*)\Psi_l$. Thus $E_H$ is minimal and
$[\B{\HH}:\R{H}(\B{\HH})]=d_\R{}^2$.

We shall show in Section~\ref{sec:CON} that $d_{\R{V}}=d_V$ (defined by
(\ref{DV})) if $\R{V}$ is a \bOG\END, \IMP able in some Fock \REP.

Let us add a last remark on the general situation. Suppose we are
given a set of \IMP ers $(\Psi_n)_{n\in I}$. Then for $m,n\in I$,
$\Psi_m\Psi_n^*\in\pi(\R{}(\frak A))'$ is a partial \ISM y containing
$\RAN\Psi_n$ in its initial space, and $\Psi_m=(\Psi_m\Psi_n^*)
\Psi_n$. This suggests to construct a complete set of \IMP ing \ISM
ies by multiplying one \ISM y $\Psi$ fulfilling (\ref{INTER}) with
certain partial \ISM ies in $\pi(\R{}(\frak A))'$. We shall employ this
idea in Section~\ref{sec:NORMFORM}.

After this digression we concentrate on \bOG\TR s again. Inspection of
(\ref{IMPQEQ}) leads us to study the \REP s $\pi_P\0\R{V}$; as will
turn out, they are \QEQ t to GNS--\REP s \ASS ed with the states
$\OM_P\0\R{V}$ (a similar observation has been made, in a different
setting, by Rideau~\cite{R}). To see this let $P$ be a \bP and
$V\in\IK{}$, and regard
\begin{equation} \label{V}
  v:=PVV^*P
\end{equation} 
as an \oP on $P(\KK)$. The direct sum \dEC $P(\KK)=\ker
v\,\2\,\4{\RAN v}$ induces a tensor product \dEC of Fock
space: $\FA{P(\KK)}\cong\FA{\ker v}\otimes\FA{\4{\RAN v}}$. Choose an
\oNB $(f_j)_{j=1,\ldots,N_V}$ for $\ker v$ where
\begin{equation} \label{NV}
  N_V:=\dim\ker v\leq\frac{1}{2}\IND V^*
\end{equation}
(the inequality follows from $\ker v\2\7\ker v\subset\ker V^*$), and
set $A(f):=a(f)\PME$ with a \UNI y $\PME$ \IMP ing $\8_{-1}$ in 
$\pi_P$ (cf.\ the proof of Lemma~\ref{lem:MIN}). Let $\IN$ be the set
of multi--indices $\3=(\3_1,\ldots,\3_l)$ as in (\ref{I}) (with finite
entries $\3_j$) and define
\begin{equation} \label{DEF}
   \begin{array}{l}
      A_\3:=A(f_{\3_1})\cdots A(f_{\3_l})\quad(A_0:=\1),\\
      \phi_\3^V:=A^*_\3\Omega_P,\\
      \FF_\3^V:=\4{\pi_P(\R{V}(\CK))\phi_\3^V},\\
      \pi_\3^V:=\pi_P\0\R{V}|_{\FF_\3^V}.
   \end{array}
\end{equation}
Note that the $A_\3$ are partial \ISM ies in $\pi_P(\R{V}(\CK))'$.
\begin{Lem}\label{lem:DEC}
  Each of the $2^{N_V}$ cyclic sub\REP s
  $(\pi_\3^V,\FF_\3^V,\phi_\3^V)$ induces the state $\OM_P\0\R{V}$,
  and $\pi_P\0\R{V}$ splits into their direct sum:\quad $\pi_P\0\R{V}=
  \bigoplus_{\3\in\IN}\pi_\3^V$.
\end{Lem}
\begin{Proofspc} 
  Invariance of $\FF_\3^V$ and cyclicity of $\phi_\3^V$ are clear by
  definition. Since $A_\3\in\pi_P(\R{V}(\CK))'$ and
  $A_\3A^*_\3\Omega_P=\Omega_P$, we have
  $\<{\phi_\3^V,\pi_\3^V(A)\phi_\3^V}=\<{\Omega_P,\pi_P(\R{V}(A))\Omega_P}
  =\OM_P(\R{V}(A)),\ A\in\CK$. Thus $(\pi_\3^V,\FF_\3^V,\phi_\3^V)$ is a
  GNS--triple for $\OM_P\0\R{V}$ (and the \REP s $\pi_\3^V$ are mutually
  \UNI ily equivalent).

  Next we show $\FF_\3^V\bot\FF_\5^V$ for $\3\neq\5$. Since at least one
  of the vectors $A_\3A^*_\5\Omega_P,\ A_\5A^*_\3\Omega_P$ vanishes if
  $\3\neq\5$, we have for $A,B\in\CK$
  $$\<{\pi_P(\R{V}(A))\phi_\3^V,\pi_P(\R{V}(B))\phi_\5^V}=
  \<{A^*_\3\Omega_P,\pi_P(\R{V}(A^*B))A^*_\5\Omega_P}=0,$$ 
  implying orthogonality of $\FF_\3^V$ and $\FF_\5^V$.

  Finally we have to prove $\FA{P(\KK)}=\bigoplus_\3\FF_\3^V$. Using
  $\pi_P(\R{V}(B(k)))=a(PVk)^*+a(PV\7k),\ k\in\KK$, one can show by
  induction on the particle number 
  $$\FF_0^V=\4{\pi_P(\R{V}(\CK))\Omega_P}=\FA{\4{\RAN PV}} =\FA{\4{\RAN
    v}}.$$
  Since the $\phi_\3^V$ form an \oNB for $\FA{\ker v}$, the
  assertion follows.
\end{Proofspc}
The \dEC of these cyclic \REP s into \IRR s will be examined
in Section~\ref{sec:DECREP}. First we state the main result of this
section. Remember that $\4{P}=\1-P$.
\begin{Th}\label{th:IMP}
  A \bOG\eND $\R{V}$ is \ISM ically \IMP able in a Fock \rEP $\pi_P$
  \iFF $PV\4{P}\in\JK{2}$.
\end{Th}
\begin{Proofspc}
  In view of (\ref{IMPQEQ}) and Lemma~\ref{lem:DEC}, $\R{V}$ is \IMP
  able in $\pi_P$ \iFF$\OM_P\0\R{V}\approx\OM_P$. Lemma~\ref{lem:OP}
  and the Powers--St\o rmer--Araki criterion in the form~(\ref{PSC})
  imply that $\OM_P\0\R{V}\approx\OM_P$ \iFF
  $\4{P}V^*PV\4{P}\in\JK{1}$. The latter condition is clearly
  equivalent to $PV\4{P}\in\JK{2}$.
%
\end{Proofspc}
Note that $PV\4{P}$ is \hS\iFF $[P,V]=PV\4{P}-\4{P}VP$ is, so the
Shale--Stinespring condition (\ref{SS}) remains valid.  We denote the
\sG of \bOG\OP s fulfilling (\ref{SS}) by
$$\IK{P}:=\{V\in\IK{}\ |\ PV\4{P}\in\JK{2}\}.$$
Since $PV\4{P}$ and $\4{P}VP$ are compact for $V\in\IK{P}$,
$(PVP)\2(\4{P}V\4{P})=V-PV\4{P}-\4{P}VP$ is semi--\FR, and
$$\IND V^*=2\IND PV^*P\in2\NN\cup\{\infty\}$$
(we used $\4{P}V\4{P}=\7(PVP)\7$). Thus we have a \dEC (cf.\
(\ref{INK}))
$$\IK{P}=\bigcup_{m\in\NN\cup\{\infty\}}\INK{2m}{P},\qquad
  \INK{2m}{P}:=\IK{P}\cap\INK{2m}{}.$$
In particular, the ``statistical \DIM'' $d_V$ defined by (\ref{DV})
is contained in $\NN\cup\{\infty\}$ if $V\in\IK{P}$. Let us finally
remark that non--surjective \bOG\END s cannot be inner since $\CK$,
being AF and thus finite, does not contain non--unitary \ISM ies.
\subsection{Equivalence of Representations}
\label{sec:DECREP}
As mentioned in Section~\ref{sec:PRE}, the $\INK{0}{}$--orbits in
$\IK{}$ w.r.t.\ left multiplication are just the subsets $\INK{n}{}$.
In the present section, we are interested in $\INK{0}{P}$--orbits (for
fixed $P$) since each such orbit gives rise to a unique equivalence
class of \REP s $\pi_P\0\R{V}$. For $V\in\IK{}$, we use the
notation
$$ S_V:=V^*PV\in\QK,\qquad Q_V:=\1-VV^*,$$
and the symbol `$\simeq$' will mean `\UNI ily equivalent'.  We only
consider the action of $\INK{0}{P}$ on the \sG of \bOG\OP s with
finite index
$$\INK{\rm fin}{}:=\{V\in\IK{}\,|\,\IND V^*<\infty\}.$$
For $V\in\INK{\rm fin}{}$, the \OP s $Q_V$ (the projection onto $\ker
V^*$) and $S_V\4{S_V}=-V^*PQ_V\4{P}V$ have finite rank.
\begin{Lem}\label{lem:EQ}
  Let $V,V'\in\INK{\rm fin}{}$. Then the following conditions are
  equivalent:\\
  a) $\pi_P\0\R{V}$ and $\pi_P\0\R{V'}$ are \UNI ily equivalent;\\
  b) there exists $U\in\INK{0}{P}$ with $V'=UV$;\\
  c) $\IND V=\IND V'$, and $S_V-S_{V'}$ is \HS.
\end{Lem}
\begin{Proofspc}
  We first show a) $\Rightarrow$ c). By Lemma~\ref{lem:DEC},
  $\pi_P\0\R{V}\simeq\pi_P\0\R{V'}$ implies $\OM_{S_V}=
  \OM_P\0\R{V}\approx\OM_P\0\R{V'}=\OM_{S_{V'}}$. Hence by
  (\ref{QEQ}), $S_V^{1/2}-S_{V'}^{1/2}\in\JK{2}$ which entails
  $S_V-S_{V'}\in\JK{2}$ \cite{PS,A70}\footnote{By an argument in
    \cite{JMB1}, the conditions $S_V^{1/2}-S_{V'}^{1/2}\in\JK{2}$ and
    $S_V-S_{V'}\in\JK{2}$ are actually equivalent for $V,V'\in\INK{\rm
      fin}{}$.}.
  Moreover, equivalent \REP s have isomorphic
  commutants. We have (cf.~\cite{A70}) $\pi_P(\R{V}(\CK))'=(\pi_P(B(\ker
  V^*))\PME)''$ with $\PME$ as in the proof of
  Lemma~\ref{lem:MIN}. Hence the commutants have \DIM s $2^{\iND V^*}$
  resp.\ $2^{\iND V'^*}$, and the indices of $V$ and $V'$ must be
  equal.

  Next we show c) $\Rightarrow$ b). Let $u$ be a partial \ISM y with
  initial space $\ker V^*$, final space $\ker V'^*$, and $u=\4{u}$
  (such $u$ exists due to $\7$--invariance and equality of \DIM s of the
  kernels). Then $U:=V'V^*+u$ is an element of $\INK{0}{}$ and
  fulfills $V'=UV$. We have to prove that $PU\4{P}\in\JK{2}$. But $u$
  has finite rank, so it suffices to show
  $$A:=\4{P}VS_{V'}V^*\4{P}\in\JK{1}.$$
  Since $S_V\4{S_V}$ and $S_{V'}\4{S_{V'}}$ have finite rank, 
  $S_{V'}\4{S_{V}}+S_{V}\4{S_{V'}}=(S_{V'}-S_V)(\4{S_V}-\4{S_{V'}})
  +S_V\4{S_V}+S_{V'}\4{S_{V'}}$ is trace class. But the same is true
  for $A=AQ_V+AVV^*=AQ_V+\4{P}V(S_{V'}\4{S_{V}}+S_{V}\4{S_{V'}})V^*+
  \4{P}Q_VPV\4{S_{V'}}V^*$.

  b) $\Rightarrow$ a) is obvious.
\end{Proofspc}
In order to make use of part c) of the lemma, we need information
about the \OP s $S_V$. An orthogonal projection $E$ on $\KK$ is called
a {\em partial \BP\/}~\cite{A70} if $E\4{E}=0$. By definition, the
$\7$--{\em co\DIM\/} of $E$ is the \dIM of $\ker(E+\4{E})$.
The following lemma holds for arbitrary $S\in\QK$ (except for the
formula for the $\7$--co\DIM, of course) as long as $S\4{S}$ has
finite rank.
\begin{Lem}\label{lem:SV}
  Let $V\in\INK{\rm fin}{}$ and let $E_V$ denote the orthogonal
  projection onto $\ker S_V\4{S_V}$. Then $S_VE_V=E_VS_V$ is a partial
  \bP with finite $\7$--co\dIM $\IND V^*-2N_V$. Moreover, there exist
  $\9_1,\ldots,\9_r\in(0,\frac{1}{2})$, partial \BP s $E_1,\ldots,E_r$
  and an orthogonal projection $E_\frac{1}{2}=\4{E_\frac{1}{2}}$ such
  that $E_V+E_\frac{1}{2}+\sum_{j=1}^r(E_j+\4{E_j})=\1$ and
  \begin{equation} \label{SV}
    S_V=S_VE_V+\frac{1}{2}E_\frac{1}{2}+
      \sum_{j=1}^r\Big(\9_jE_j+(1-\9_j)\4{E_j}\Big).
  \end{equation}
\end{Lem}
\begin{Proofspc}
  Since $S_V\4{S_V}=S_V-S_V^2$, $S_V$ commutes with $E_V$ and fulfills
  $S_VE_V=S_V^2E_V$ and also $(S_VE_V)(\7S_VE_V\7)=S_V\4{S_V}E_V=0$.
  Hence $S_VE_V$ is a partial \BP. The \dIM of
  $\ker(S_VE_V+\4{S_VE_V}) =\ker E_V$ (the $\7$--co\dIM of
  $S_VE_V$) equals the rank of $S_V\4{S_V}$ which is finite for
  $V\in\INK{\rm fin}{}$. By $S_V\4{S_V}=V^*PQ_VPV$, the rank of
  $S_V\4{S_V}$ equals $\dim V^*P(\ker V^*)$. Now consider the
  decomposition
  $$\ker V^*= \ker v\2\ker\4{v}\2\Big(\ker V^*\ominus(\ker
  v\2\ker\4{v})\Big)$$ with $v$ given by (\ref{V}). $V^*P$ vanishes on
  $\ker v\2\ker\4{v}$, but the restriction of $V^*P$ to $\ker
  V^*\ominus(\ker v\2\ker\4{v})$ is one--to--one since $V^*Pk=0=V^*k$
  implies $V^*\4{P}k=0$, i.e.\ $k\in\ker v\2\ker\4{v}$. Hence the
  $\7$--co\dIM of $S_VE_V$ equals $\dim(\ker V^*\ominus(\ker
  v\2\ker\4{v}))=\IND V^*-2N_V$.

  Let $s_V$ denote the restriction of $S_V$ to $\RAN S_V\4{S_V}$.
  $s_V$ is a positive \oP on a finite \DIM al \hSP and has a complete
  set of eigenvectors with eigenvalues in $(0,1)$. If $\9$ is an
  eigenvalue of $s_V$, then $1-\9$ is also an eigenvalue (with the
  same multiplicity) due to $s_V+\4{s_V}=\1-E_V$. Thus there exist
  $\9_1,\ldots,\9_r\in(0,\frac{1}{2})$ and spectral projections
  $E_\frac{1}{2},E_1,\ldots,E_r$ with
  $\4{E_\frac{1}{2}}=E_\frac{1}{2},\ E_j\4{E_j}=0$ such that
  $E_\frac{1}{2}+\sum_{j=1}^r(E_j+\4{E_j})=\1-E_V$ and
  $s_V=\frac{1}{2}E_\frac{1}{2}+\sum_{j=1}^r(\9_jE_j+(1-\9_j)\4{E_j})$.
\end{Proofspc}
As a consequence, \OP s $S_V$ with $\IND V^*=1$ necessarily have the
form $S_V=S_VE_V+\frac{1}{2}E_\frac{1}{2}$ where
$E_\frac{1}{2}=\1-E_V$ has rank one.  By taking direct sums of
$V\in\INK{1}{}$ with \OP s $V(\PHI)$ from the example below, we see
that each combination of eigenvalues and multiplicities that is
allowed by Lemma~\ref{lem:SV} actually occurs for some $S_{V'}$. We
further remark that a \qS $\OM_S$ with $S$ of the form (\ref{SV}) is a
product state\footnote{A state $\OM$ is a {\em product state\/}
  w.r.t.\ a \dEC $\KK=\2_j\KK_j$ of $\KK$ into closed, $\7$--\iNV
  subspaces if $\OM(AB)=\OM(A)\OM(B)$ whenever $A\in{\cal
    C}(\KK_j,\7|_{\KK_j}),\ B\in{\cal
    C}(\KK_j^\bot,\7|_{\KK_j^\bot})$. In this case, the restrictions
  $\OM_j$ of $\OM$ to ${\cal C}(\KK_j,\7|_{\KK_j})$ are {\em even\/}
  (i.e.~\iNV under $\8_{-1}$) with at most one exception. If all
  $\OM_j$ are even, then $\OM$ is pure \iFF all $\OM_j$ are
  \cite{P67}.}
as defined by Powers~\cite{P67} (see also~\cite{MRT,Man}) w.r.t.\ the
\dEC $\KK=\ker S\4{S}\2\RAN
E_\frac{1}{2}\2\bigoplus_j\RAN(E_j+\4{E_j})$.  Clearly, the
restriction of $\OM_S$ to ${\cal C}(\ker S\4{S},\7|_{\ker S\4{S}})$ is
a Fock state, the restriction to ${\cal C}(\RAN E_\frac{1}{2},\7|_{\RAN
  E_\frac{1}{2}})$ the central state.
\\[1.5ex]{\it Example}. Let $(f_n)_{n\in\NN}$ be an \oNB for
  $P(\KK)$ and set $E_n:=f_n\<{f_n,.\,},\
  f_n^+:=(f_n+\7f_n)/\sqrt{2},\ f_n^-:=i(f_n-\7f_n)/\sqrt{2}$. Then
  $(f_n^s)_{s=\pm,\,n\in\NN}$ is an \oNB for $\KK$
  consisting of $\7$--\iNV vectors. For $\PHI\in\RR$, define a \bOG\oP
  $$V(\PHI):=(f_0^+\cos\PHI+f_1^-\sin\PHI)\<{f_0^+,.\,}+
    (f_0^-\sin\PHI-f_1^+\cos\PHI)\<{f_0^-,.\,}+
    \sum_{s=\pm,\,n\geq1}f_{n+1}^s\<{f_n^s,.\,}.$$
  Then $V(\PHI)\in\INK{2}{}$, and the eigenvalue
  $\9_\PHI=(1+\sin2\PHI)/2$ of $S_{V(\PHI)}=\Big(\9_\PHI
  E_0+(1-\9_\PHI)\4{E_0}\Big)+\sum_{n\geq1}E_n$ assumes any value in
  $[0,1]$ as $\PHI$ varies over $[-\pi/4,\pi/4]$.\\[1.5ex]
Next we characterize the \bOG\OP s $V$ for which $S_V$ takes a
particularly simple form. A distinction arises between the cases of
even and odd \fR index.
\begin{Lem}\label{lem:PURE}
  a) Let $W\in\IK{}$. Then the following conditions are equivalent:\\
  \hspace*{\parindent}(i) $\OM_P\0\R{W}$ is a pure state;\\
  \hspace*{\parindent}(ii) $S_W$ is a \BP;\\
  \hspace*{\parindent}(iii) $[P,WW^*]=0$.\\
  If any of these conditions is fulfilled, then $\IND W^*=2N_W$ and
  $\pi_P\0\R{W}\simeq d_W\cdot\pi_{S_W}$.\\ 
  b) For any \bP $P'$ and $m\in\NN\cup\{\infty\}$, there exists
  $W\in\INK{2m}{}$ with $S_W=P'$.\\
  c) Let $W\in\INK{\rm fin}{}$. Then the following conditions are
  equivalent:\\ 
  \hspace*{\parindent}(i) $\OM_P\0\R{W}$ is a mixture of two disjoint
    pure states;\\ 
  \hspace*{\parindent}(ii) $S_WE_W$ is a partial \bP with $\7$--co\dIM 1;\\
  \hspace*{\parindent}(iii) $[P,WW^*]$ has rank 2;\\
  \hspace*{\parindent}(iv) $\IND W^*=2N_W+1$.\\
  d) For any partial \bP $P'$ with $\7$--co\dIM 1 and
  $m\in\NN\cup\{\infty\}$, there exists $W\in\INK{2m+1}{}$ with
  $S_WE_W=P'$.
\end{Lem}
\begin{Proofspc}
  a) We know from Section~\ref{sec:PRE} that $\OM_P\0\R{W}$ is pure
  \iFF $S_W$ is a projection. We have
  $$S_W^2=S_W\iff W^*PQ_WPW=0\iff Q_WPWW^*=0\iff[P,WW^*]=0.$$
  If this is fulfilled, $\ker WW^*=\ker(PWW^*P)\2\ker(\4{P}WW^*\4{P})$
  has \dIM $2N_W$. By Lemma~\ref{lem:DEC}, $\pi_P\0\R{W}$ is the
  direct sum of $d_W=2^{N_W}$ \iRR sub\REP s, each equivalent to the
  Fock \rEP $\pi_{S_W}$.

  b) Let $m$ and $P'$ be given. There clearly exists $W'\in\INK{2m}{}$ with
  $[P,W']=0$. Since $\INK{0}{}$ acts transitively
  on the set of \BP s, we may choose $U\in\INK{0}{}$ with
  $U^*PU=P'$. Then $W:=W'U$ has the desired properties.

  c) (ii) $\IFFF$ (iii) follows from the facts that the $\7$--co\dIM
  of $S_WE_W$ equals the rank of $WW^*PQ_W$ (cf.\ the proof of
  Lemma~\ref{lem:SV}) and that $[P,WW^*]=Q_WPWW^*-WW^*PQ_W$. (ii) and
  (iv) are equivalent by virtue of Lemma~\ref{lem:SV}. (ii)
  $\Rightarrow$ (i) has been shown by Araki~\cite{A70}. To prove (i)
  $\Rightarrow$ (iv), assume that $\IND W^*-2N_W>1$ (if $\IND
  W^*=2N_W$, then $S_W$ is a \bP and $\OM_P\0\R{W}$ pure). By
  Lemma~\ref{lem:SV}, there exist a two--\DIM al, $\7$--\iNV subspace
  $\KK_1\subset\KK$, $\9\in(0,1)$ and a \bP $E$ of $(\KK_1,\7_1)$,
  $\7_1:=\7|_{\KK_1}$, such that $S_1:=S|_{\KK_1}=\9E+(1-\9)\4{E}$.
  Set $\KK_2:=\KK_1^\bot,\ \7_2:=\7|_{\KK_2}$ and $S_2:=S|_{\KK_2}$.
  Then $\OM_S$ is a product state w.r.t.\ $\KK=\KK_1\2\KK_2$; we write
  $\OM_S=\OM_{S_1}\otimes\OM_{S_2}$ which means that
  $\OM_S(A_1A_2)=\OM_{S_1}(A_1)\OM_{S_2}(A_2),\ A_j\in\CKJ{j}$. As
  $\OM_{S_1}=\9\OM_E+(1-\9)\OM_\4{E}$ is a mixture of two equivalent
  Fock states over $\CKJ{1}$,
  $\OM_S=\9(\OM_E\otimes\OM_{S_2})+(1-\9)(\OM_\4{E}\otimes\OM_{S_2})
  =\9\OM_{E+S_2}+(1-\9)\OM_{\4{E}+S_2}$ is a mixture of two \QEQ t \QS
  s over $\CK$. We are going to show that $\OM_E\otimes\OM_{S_2}$ and
  $\OM_\4{E}\otimes\OM_{S_2}$ are orthogonal. It is readily seen that
  $\OM_{S_1}$ is faithful. Let $(\pi_1,\HH_1,\Omega_1)$ be the
  GNS--\rEP for $\OM_{S_1}$, i.e.\ $\HH_1=\CKJ{1}$ as a vector space,
  $\Omega_1=\1$, $\pi_1$ acts by left multiplication, and
  $\OM_{S_1}(A)=\<{\Omega_1,\pi_1(A)\Omega_1}$. Since $\OM_{S_1}$ is
  even, we may \iMP $\8_{-1}$ by the self--adjoint
  \UNI y $\PEME$ with
  $\PEME\pi_1(A)\Omega_1=\pi_1(\8_{-1}(A))\Omega_1$. Now choose a
  unit vector $e\in E(\KK_1)$, set $e^+:=(e+\7e)/\sqrt{2},\ 
  e^-:=i(e-\7e)/\sqrt{2}$, and define complementary orthogonal
  projections $P^\pm:=\frac{1}{2}\1\pm
  i\pi_1(B(e^-)B(e^+))\PEME\in\pi_1(\CKJ{1})'$. A computation shows
  $$\<{\Omega_1,\pi_1(A)P^+\Omega_1}=\9\OM_E(A),\qquad
    \<{\Omega_1,\pi_1(A)P^-\Omega_1}=(1-\9)\OM_\4{E}(A),\qquad
    A\in\CKJ{1}.$$
  Let $(\pi_2,\HH_2,\Omega_2)$ be the GNS--\rEP for $\OM_{S_2}$. Then
  the GNS--\rEP $(\pi_S,\HH_S,\Omega_S)$ for $\OM_S$ may be identified
  with the $\ZZ_2$--graded tensor product of $\pi_1$ and $\pi_2$.
  Since $\deg P^\pm=0$ and $P^\pm\in\pi_1(\CKJ{1})'$, the projections
  $P^\pm\otimes\1$ lie in the commutant $\pi_S(\CK)'$. We now infer
  from $\<{\Omega_S,\pi_S(A)(P^+\otimes\1)\Omega_S}=
  \9(\OM_E\otimes\OM_{S_2})(A)$ and
  $\<{\Omega_S,\pi_S(A)(P^-\otimes\1)\Omega_S}=
  (1-\9)(\OM_\4{E}\otimes\OM_{S_2})(A)$ that $\OM_E\otimes\OM_{S_2}$
  and $\OM_\4{E}\otimes\OM_{S_2}$ are indeed orthogonal. Hence $\OM_S$
  cannot be a mixture of two disjoint pure states. This proves (i)
  $\Rightarrow$ (iv) and therefore part c).

  d) Let $(f_n)_{n\in\NN}$ be an \oNB for $P(\KK)$, $(g_n)_{n\geq1}$
  an \oNB for $P'(\KK)$, and $g_0$ a unit vector in
  $\ker(P'+\4{P'})$. Set
  $V:=f_0^+\<{g_0,.\,}+\sum_{s=\pm,\,n\geq1}f_n^s\<{g_n^s,.\,}$ (we
  use the notation of the example). Then $V\in\INK{1}{}$ and
  $S_V=\frac{1}{2}g_0\<{g_0,.\,}+P'$. This implies $S_VE_V=P'$, and if
  we choose $W'$ as in the proof of b), then $W:=W'V$ has the desired
  properties.
\end{Proofspc}
One may use the argument given in the proof of c) inductively to show
that a \qS $\OM_S$ with $S$ of the form (\ref{SV}) is a mixture of
$2^m$ mutually orthogonal, pure states if the rank of $S\4{S}$ is $2m$
or $2m-1$.

Now let us discuss the \dEC of \REP s $\pi_P\0\R{V}$ with
$V\in\INK{\rm fin}{}$. If $\IND V^*$ is even (resp.\ odd), then
$S_VE_V$ is a partial \bP with even (odd) $\7$--co\dIM by
Lemma~\ref{lem:SV}, and there exists a \bP (partial \bP with
$\7$--co\dIM 1) $P'$ with $P'-S_V\in\JK{2}$ (we may choose $P'$ to
coincide with $S_VE_V$ on $\ker S_V\4{S_V}$; then $P'-S_V$ has finite
rank). By Lemma~\ref{lem:PURE}, there exists $W$ with $\IND W=\IND V$
and $S_WE_W=P'$, and Lemma~\ref{lem:EQ} implies
$\pi_P\0\R{V}\simeq\pi_P\0\R{W}$. The latter \rEP splits into
$2^{N_W}$ copies of the GNS--\rEP $\pi_{S_W}$ for the state
$\OM_P\0\R{W}$ by Lemma~\ref{lem:DEC}. If $\IND V^*$ is even,
$\pi_{S_W}=\pi_{P'}$ and $2^{N_W}=d_V$. If $\IND V^*$ is odd, then
$\pi_{S_W}=\pi^+\2\pi^-$ where $\pi^\pm$ are mutually inequivalent,
\IRR, so--called {\em pseudo Fock \REP s\/} by virtue of a lemma of
Araki (see~\cite{A70} for details), and $2^{N_W}=2^{-1/2}d_V$.

Summarizing, we rediscover B\"ockenhauer's result~\cite{JMB2}:
\begin{Th}\label{th:DEC}
  Let $P$ be a \bP and $V\in\INK{\rm fin}{}$. If $\IND V^*$ is even,
  then there exist \BP s $P'$ with $P'-S_V\in\JK{2}$, and for each
  such $P'$
  $$\pi_P\0\R{V}\simeq d_V\cdot\pi_{P'}.$$
  If $\IND V^*$ is odd, there exist partial \BP s $P'$ with
  $\7$--co\dIM 1 and $P'-S_V\in\JK{2}$. For each such $P'$,
  $$\pi_P\0\R{V}\simeq 2^{-1/2}d_V\cdot(\pi_{P'}^+\2\pi_{P'}^-)$$
  where $\pi_{P'}^\pm$ are the (inequivalent, \IRR) pseudo Fock \REP s
  induced by $P'$.
\end{Th}

We shall study the action of $\INK{0}{P}$ on $\IK{P}$ in
Section~\ref{sec:DEC}. Then the orbits are the sets $\INK{2m}{P},\
m\leq\infty$, and each orbit contains representatives $W$ with
$[P,WW^*]=0$. 
\section{Construction of Implementing Isometries}
\label{sec:CON}
Our construction of \IMP ers follows the lines of \RUI' approach
in~\cite{R78} which is to our knowledge the most complete treatment of
the \IMP ation of \bOG\AUT s. Another advantage of~\cite{R78} for our
purposes is the (implicit) use of Araki's \sDC formalism.

Let us first introduce some notation, followed by simple
observations. Throughout this section $P_1$ is a fixed
\bP of $(\KK,\7)$ and $P_2:=\1-P_1=\4{P_1}$. The components of an \oP
$A$ on $\KK$ are denoted by
$$A_{mn}:=P_mAP_n,\quad m,n=1,2$$
and are regarded as \OP s from $\KK_n:=P_n(\KK)$ to $\KK_m$. Thus $\ker
A_{mn}$, $(\ker A_{mn})^\bot$, $(\RAN A_{nm})^\bot$ etc.\ are viewed
as subspaces of $\KK_n$, and we have
$${A_{mn}}^*={A^*}_{nm},\quad\4{A_{11}}=\4{A}_{22}\quad\mbox{etc.}$$
We also use matrix notation $A=\mAT{A_{11}}{A_{12}}{A_{21}}{A_{22}}$
w.r.t.\ to the \dEC $\KK=\KK_1\2\KK_2$.

Let $V\in\IK{P_1}$ be a fixed \bOG\OP, with $\R{V}$ \IMP able in the
Fock \rEP $\pi_{P_1}$. The relation $V^*V=\1$ reads in components
\begin{eqnarray}
   {V_{11}}^*V_{11}+{V_{21}}^*V_{21} &=& P_1, \label{REL1} \\
   {V_{22}}^*V_{22}+{V_{12}}^*V_{12} &=& P_2, \label{REL2} \\
   {V_{11}}^*V_{12}+{V_{21}}^*V_{22} &=& 0,   \label{REL3} \\
   {V_{22}}^*V_{21}+{V_{12}}^*V_{11} &=& 0,   \label{REL4}
\end{eqnarray}
whereas $V=\4{V}$ gives
$$V_{22}=\4{V_{11}},\quad V_{21}=\4{V_{12}}.$$
Since $V_{12}$ is a \hS\oP by Theorem~\ref{th:IMP},
${V_{22}}^*V_{22}$ is \fR (with vanishing index) by (\ref{REL2}). This
means in particular 
\begin{equation} \label{LVFIN}
  \dim\ker V_{22}=\dim\ker{V_{22}}^*V_{22}<\infty.
\end{equation}
Note that $V_{12}|_{\ker V_{22}}$ is \ISM ic and, by (\ref{REL3}), 
\begin{equation} \label{KER}
  V_{12}(\ker V_{22})\subset\ker{V_{11}}^*.
\end{equation}
As mentioned at the end of Section~\ref{sec:IMP}, ${V_{11}}^*$ is
semi--\fR with $\IND{V_{11}}^*=\frac{1}{2}\IND V^*$. By the above and
by $\ker V_{11}=\7\ker V_{22}$, we have
\begin{equation} \label{DIMKER}
  \IND{V_{11}}^*=\dim(\ker{V_{11}}^*\ominus V_{12}(\ker V_{22})).
\end{equation}

In the following, we are going to describe some \OP s by integral
kernels. Thus we assume in this section (without loss of generality)
$$\KK_1=L^2(\RR^d).$$
\subsection{Unbounded Bilinear Hamiltonians and \RUI' Operator $\LL{}$}
\label{sec:BH}
Bounded {\em\BH s\/} have been introduced by Araki~\cite{A68} as
infinitesimal generators of one--parameter groups of inner \bOG\AUT s.
More specifically, one may assign to a finite rank \oP
$H=\sum_jk_j\<{k_j',.\,}$ on $\KK$ the \bH
$$b(H):=\sum_jB(k_j)B(k_j')^*$$
and extend $b$ to a linear map from $\JK{1}$ to $\CK$ by continuity
(relative to trace norm on $\JK{1}$ and C*--norm on $\CK$). If $H\in\JK{1}$
satisfies $H^*=-H$ and $\4{H}=H$, then $b(H)/2$ is the generator of
the one--parameter group $(\8_{e^{tH}})_{t\in\RR}$:
$$\8_{e^{tH}}=\AD(\exp(tb(H)/2)).$$
Further properties of $b$ are summarized in~\cite{A70,A87}.

Since elements $B(k)$ with $k\in\KK_1$ correspond to creation \OP s in
the Fock \rEP $\pi_{P_1}$, we may write
$$\pi_{P_1}(b(H))=H_{12}a^*a^*+H_{11}a^*a+H_{22}aa^*+H_{21}aa$$
where the terms on the right are defined by
$H_{12}a^*a^*:=\pi_{P_1}(b(H_{12}))$ etc. Introducing {\em Wick
 ordering\/} by $\WO{a(f)a(g)^*} = -a(g)^*a(f)$, we get
$\WO{H_{22}aa^*} =-\4{{H_{22}}^*}a^*a=H_{22}aa^*-(\TRA H_{22})\1$ and
\begin{equation} \label{WICK1}
  \WO{\pi_{P_1}(b(H))}=H_{12}a^*a^*+(H_{11}-\4{{H_{22}}^*})a^*a+H_{21}aa.
\end{equation}

According to~\cite{R78,CR}, one may define such Wick ordered
expressions for {\em bounded\/} $H$ as follows. Let
$\SSS\subset\FA{\KK_1}$ be the dense subspace consisting of finite
particle vectors $\phi$ with $n$--particle wave functions $\phi^{(n)}$
in the Schwartz space $\SSS(\RR^{dn})$. For $p\in\RR^d$, the unsmeared
\aNN\oP $a(p)$ with (invariant) domain $\SSS$ is defined by
$$(a(p)\phi)^{(n)}(p_1,\ldots,p_n):=\sqrt{n+1}\phi^{(n+1)}
  (p,p_1,\ldots,p_n).$$
Since $a(p)$ is not closable, one defines $a(p)^*$ as the \qF adjoint
of $a(p)$ on $\SSS\times\SSS$. Then Wick ordered monomials
$a(q_m)^*\cdots a(q_1)^*a(p_1)\cdots a(p_n)$ are well--defined \QF s on
$\SSS\times\SSS$, and for $\phi,\phi'\in\SSS$, 
$$\<{\phi,a(q_m)^*\cdots a(q_1)^*a(p_1)\cdots a(p_n)\phi'}
  =\<{a(q_1)\cdots a(q_m)\phi,a(p_1)\cdots a(p_n)\phi'}$$ 
is a function in $\SSS(\RR^{d(m+n)})$ to which tempered
distributions may be applied. For example, one has in the \qF sense
$$\begin{array}{ccc}
  a(f)=\int\4{f(p)}a(p)\,dp, & a(f)^*=\int f(p)a(p)^*\,dp, & f\in\KK_1.
\end{array}$$

Now let $H$ be a bounded \oP on $\KK$. By the nuclear theorem of
Schwartz, there exist tempered distributions $H_{mn}(p,q),\ m,n=1,2$,
given by
\begin{equation} \label{WICK2}
   \begin{array}{rcl}
      \<{f,H_{11}g} &=& \int\4{f(p)}H_{11}(p,q)g(q)\,dp\,dq, \\
      \<{f,H_{12}\7g} &=& \int\4{f(p)}H_{12}(p,q)\4{g(q)}\,dp\,dq, \\
      \<{\7f,H_{21}g} &=& \int f(p)H_{21}(p,q)g(q)\,dp\,dq, \\
      \<{\7f,H_{22}\7g} &=& \int f(p)H_{22}(p,q)\4{g(q)}\,dp\,dq, 
         \quad f,g\in\SSS(\RR^d)\subset\KK_1.
   \end{array}
\end{equation}
Hence we may define the following \QF s on $\SSS\times\SSS$
\begin{equation} \label{WICK3}
   \begin{array}{rcl}
      H_{12}a^*a^* &:=& \int H_{12}(p,q)a(p)^*a(q)^*\,dp\,dq \\
      H_{11}a^*a &:=& \int H_{11}(p,q)a(p)^*a(q)\,dp\,dq \\
      :\!H_{22}aa^*\!: &:=& -\4{{H_{22}}^*}a^*a \\
        &=& -\int H_{22}(q,p)a(p)^*a(q)\,dp\,dq \\
      H_{21}aa &:=& \int H_{21}(p,q)a(p)a(q)\,dp\,dq.
   \end{array}
\end{equation}
The (Wick ordered, unbounded) \bH induced by $H$ is then defined in
analogy to (\ref{WICK1}) as
$$\WO{b(H)} :=H_{12}a^*a^*+(H_{11}-\4{{H_{22}}^*})a^*a+H_{21}aa\,;$$
it is linear in $H$. We define its Wick ordered powers as
\begin{equation} \label{WICK4}
   \WO{b(H)^l} :=l!\sum_{l_1,l_2,l_3=0\atop l_1+l_2+l_3=l}^l
   \frac{1}{l_1!l_2!l_3!}
   (H_{12})^{l_1}(H_{11}-\4{{H_{22}}^*})^{l_2}(H_{21})^{l_3}
   a^{*2l_1+l_2}a^{l_2+2l_3}
\end{equation}
where the terms on the right hand side are \QF s on $\SSS\times\SSS$
(cf.~\cite{R78})
\begin{eqnarray}
   \lefteqn{(H_{12})^{l_1}(H_{11}-\4{{H_{22}}^*})^{l_2}(H_{21})^{l_3}
      a^{*2l_1+l_2}a^{l_2+2l_3} :=} \nonumber\\
   & & \int H_{12}(p_1,q_1)\cdots H_{12}(p_{l_1},q_{l_1})
      (H_{11}(p_1',q_1')-H_{22}(q_1',p_1'))\cdots 
      (H_{11}(p_{l_2}',q_{l_2}')-H_{22}(q_{l_2}',p_{l_2}'))
      \cdot\label{MONOM}\\
   & & {}\cdot H_{21}(p_1'',q_1'')\cdots H_{21}(p_{l_3}'',q_{l_3}'')
      a(p_1)^*\cdots a(p_{l_1})^*a(q_{l_1})^*\cdots a(q_1)^*
      a(p_1')^*\cdots a(p_{l_2}')^* a(q_{l_2}')\cdots a(q_1')
      \cdot\nonumber\\
   & & {}\cdot a(p_1'')\cdots a(p_{l_3}'') a(q_{l_3}'')\cdots a(q_1'')
      \,dp_1\,dq_1\ldots dp_{l_1}\,dq_{l_1}\,dp_1'\,dq_1'\ldots
      dp_{l_2}'\,dq_{l_2}'\,dp_1''\,dq_1''\ldots dp_{l_3}''\,dq_{l_3}''.
      \nonumber
\end{eqnarray}
Finally, we define the Wick ordered exponential
\begin{equation} \label{WICK5}
   \EH:=\sum_{l=0}^\infty
   \frac{1}{l!2^l}\WO{b(H)^l}
\end{equation}
which is also a well--defined \qF on $\SSS\times\SSS$ since the sum in
(\ref{WICK5}) is finite when applied to vectors $\phi,\phi'\in\SSS$.

How do we have to choose $H$ in order to relate these \QF s to \IMP
ers for $\R{V}$? Let us first remark that we may restrict attention to
{\em \AS\/} $H$, i.e.\ to \OP s fulfilling
\begin{equation} \label{AS}
  H=-\4{H^*}
\end{equation}
(in components: $H_{11}=-\4{{H_{22}}^*},\ H_{12}=-\4{{H_{12}}^*},\
H_{21}=-\4{{H_{21}}^*}$). Indeed, we have $H=H^++H^-$ with
$H^\pm:=\frac{1}{2}(H\pm\4{H^*})=\pm\4{(H^\pm)^*}$, and we claim that
$$\WO{b(H)}=\ \WO{b(H^-)}$$
or equivalently $\WO{b(H^+)}=0$. From $\4{(H^+_{22})^*}=H^+_{11}$
we infer $\WO{b(H^+)}=H^+_{12}a^*a^*+H^+_{21}aa$. It follows from
$H^+_{12}=\4{(H^+_{12})^*}$ and (\ref{WICK2}) that
$H_{12}(p,q)=H_{12}(q,p)$. But by virtue of the CAR we have
$H^+_{12}a^*a^*=\int H_{12}(p,q)a(p)^*a(q)^*\,dp\,dq=-\int
H_{12}(q,p)a(q)^*a(p)^*\,dp\,dq=-H^+_{12}a^*a^*$, hence
$H^+_{12}a^*a^*=0$. A similar argument shows $H^+_{21}aa=0$ which
proves the assertion.

So let $H$ be \aS in the above sense. Of course, we would like to deal
with well--defined \OP s instead of \QF s. By a result of
\RUI~\cite{R78}, $\EH$ is the \qF of a densely defined \oP with domain
$\DD:=\pi_{P_1}({\cal C}_0(\KK,\7)) \Omega_{P_1}$, the subspace of
algebraic tensors in $\FA{\KK_1}$, {\em provided that $H_{12}$ is
  \HS}. In this case, the series (\ref{WICK5}) converges strongly on
$\DD$, $\EH$ (viewed as an \OP) maps $\DD$ into the dense subspace of
$C^\infty$--vectors for the number \OP, and
\begin{equation} \label{OMNORM}
  \|\EH\Omega_{P_1}\|=(\det(P_1+H_{12}{H_{12}}^*))^{1/4}.
\end{equation}
The \OP s $H$ of interest are now selected by intertwining
properties (cf.~(\ref{INTER})) of $\EH$. Let
$a_V(f)=a(V_{11}f)+a(V_{12}\7f)^*$ denote the transformed \aNN\oP as
in (\ref{TR}). We are looking for \OP s $H$ fulfilling
\begin{eqnarray}
  \EH a(f)^* &=& a_V(f)^*\EH,\quad f\in\KK_1\label{INTER1} \\
  \EH a(g) &=& a_V(g)\EH,\quad g\in(\ker V_{11})^\bot\label{INTER2}
\end{eqnarray}
on $\DD$. Since $a_V(g)$ is a creation \oP for $g\in\ker V_{11}$,
(\ref{INTER2}) cannot hold for such $g$ unless $g=0$ (the l.h.s.\ of
(\ref{INTER2}) vanishes on $\Omega_{P_1}$, but the r.h.s.\ does not,
cf.\ the proof of Lemma~\ref{lem:L}).
We impose an additional relation for vectors in $\ker V_{11}$ which
will prove to be ``correct'':
\begin{equation} \label{INTER3}
  \EH a(h)^*=0,\quad h\in\ker V_{11}.
\end{equation}
To solve (\ref{INTER1})--(\ref{INTER3}), we have to compute
commutation relations of $\EH$ with creation
and \aNN\OP s.
\begin{Lem} \label{lem:REL}
  Let $H\in\B{\KK}$ be \aS in the sense of (\ref{AS}) with $H_{12}$
  \HS. For $f,g\in\KK_1$, the following relations hold on $\DD$ 
  \begin{eqnarray*}
    [\EH,a(f)^*] &=& a(H_{11}f)^* 
      \EH+\EH a(\7H_{21}f),\\{}
    [\EH,a(g)] &=& a(H_{12}\7g)^*
      \EH+\EH a(\4{H_{22}}g).
  \end{eqnarray*}
\end{Lem}
\begin{Proof}
  Let us first compute commutation relations for Wick monomials of the
  form (cf.~(\ref{MONOM}))
  $$H_{l_1,l_2,l_3}:=(H_{12})^{l_1}(H_{11})^{l_2}(H_{21})^{l_3}
    a^{*2l_1+l_2}a^{l_2+2l_3}.$$
  Using the formal CAR, we get
  $$\begin{array}{l}
    a(q_l)\cdots a(q_1)a(p)^* = (-1)^la(p)^*a(q_l)\cdots a(q_1)
      +\sum_{j=1}^l(-1)^{j-1}\delta(q_j-p)a(q_l)
      \cdots\WT{a(q_j)}\cdots a(q_1), \\
    (-1)^la(p_1)\cdots a(p_l)a(p)^* = a(p)^*a(p_1)\cdots a(p_l)
      +\sum_{j=1}^l(-1)^j\delta(p_j-p)a(p_1)
      \cdots\WT{a(p_j)}\cdots a(p_l),
  \end{array}$$
  where the factors under the symbol ``$\WT{\quad}$'' are to be
  omitted. In the following computation, we use in addition 
  $$\begin{array}{rclcl}
    \int a(p)H_{21}(p,q)f(q)\,dp\,dq &=& a(\7H_{21}f), && \\
    \int f(p)H_{21}(p,q)a(q)\,dp\,dq &=& a({H_{21}}^*\7f) &=& -a(\7H_{21}f),\\
    \int a(p)^*H_{11}(p,q)f(q)\,dp\,dq &=& a(H_{11}f)^*. &&
  \end{array}$$
  \begin{eqnarray*}
      \lefteqn{H_{l_1,l_2,l_3}a(f)^*=\int H_{12}(p_1,q_1)\cdots
         H_{12}(p_{l_1},q_{l_1})H_{11}(p_1',q_1')\cdots H_{11}
         (p_{l_2}',q_{l_2}') H_{21}(p_1'',q_1'')
         \cdots H_{21}(p_{l_3}'',q_{l_3}'')\ \cdot} \\
      && {}\cdot f(p)a(p_1)^*\cdots a(p_{l_1})^*a(q_{l_1})^*\cdots
         a(q_1)^* a(p_1')^*\cdots a(p_{l_2}')^*a(q_{l_2}')\cdots a(q_1')
         a(p_1'')\cdots a(p_{l_3}'')\cdot \\
      &&{}\cdot a(q_{l_3}'')\cdots a(q_1'')a(p)^*\,dp_1\,dq_1\ldots
         dp_{l_1}\,dq_{l_1}\,dp_1'\,dq_1' \ldots dp_{l_2}'\,dq_{l_2}'
         \,dp_1''\,dq_1''\ldots dp_{l_3}''\, dq_{l_3}''\,dp \\
      &=&(-1)^{l_3}\int H_{12}(p_1,q_1)\cdots
         H_{21}(p_{l_3}'',q_{l_3}'')f(p)a(p_1)^*\cdots a(q_1')
         a(p_1'')\cdots a(p_{l_3}'')a(p)^*a(q_{l_3}'')
         \cdots a(q_1'')\,dp_1\ldots dp \\
      && {}+\sum_{j=1}^{l_3} (-1)^{j-1} \int H_{12}(p_1,q_1)
         \cdots H_{21}(p_j'',q_j'')\cdots
         H_{21}(p_{l_3}'',q_{l_3}'')\delta(q_j''-p)f(p)\cdot \\
      && {}\cdot a(p_1)^*\cdots a(q_1') a(p_1'')\cdots 
         \WT{a(p_j'')} \cdots a(p_{l_3}'')
         a(q_{l_3}'')\cdots \WT{a(q_j'')} \cdots
         a(q_1'')(-1)^{l_3-j+l_3-1}a(p_j'')\,dp_1\ldots dp \\
      &=&\int H_{12}(p_1,q_1)\cdots H_{21}(p_{l_3}'',q_{l_3}'')
         f(p)a(p_1)^*\cdots a(p_{l_2}')^* a(q_{l_2}')\cdots 
         a(q_1')a(p)^* a(p_1'')\cdots a(q_1'')\,dp_1\ldots dp \\
      && {}+\sum_{j=1}^{l_3} (-1)^j \int H_{12}(p_1,q_1)
         \cdots H_{21}(p_j'',q_j'')\cdots H_{21}(p_{l_3}'',q_{l_3}'')
         \delta(p_j''-p)f(p)a(p_1)^*\cdots a(q_1') a(p_1'')\cdots \\
      && \cdots \WT{a(p_j'')} \cdots a(p_{l_3}'')
         a(q_{l_3}'')\cdots \WT{a(q_j'')} \cdots
         a(q_1'')(-1)^{j-1}a(q_j'')\,dp_1\ldots dp
         +l_3H_{l_1,l_2,l_3-1}a(\7H_{21}f) \\
      &=&(-1)^{l_2}\int H_{12}(p_1,q_1)\cdots H_{21}(p_{l_3}'',q_{l_3}'')
         f(p) a(p_1)^*\cdots a(p_{l_2}')^*a(p)^*a(q_{l_2}')
         \cdots a(q_1'')\,dp_1\ldots dp \\
      && {}+\sum_{j=1}^{l_2} (-1)^{j-1} \int H_{12}(p_1,q_1)\cdots
         H_{11}(p_j',q_j')\cdots H_{21}(p_{l_3}'',q_{l_3}'')
         \delta(q_j'-p)f(p)(-1)^{j-1}a(p_j')^* \cdot \\
      && {}\cdot a(p_1)^*\cdots a(q_1)^* a(p_1')^*\cdots
         \WT{a(p_j')^*}\cdots a(p_{l_2}')^*
         a(q_{l_2}')\cdots \WT{a(q_j')}\cdots
         a(q_1')a(p_1'')\cdots a(q_1'')\,dp_1\ldots dp \\
      && {}+2l_3H_{l_1,l_2,l_3-1}a(\7H_{21}f) \\
      &=&a(f)^*H_{l_1,l_2,l_3}+l_2a(H_{11}f)^*H_{l_1,l_2-1,l_3}
         +2l_3H_{l_1,l_2,l_3-1}a(\7H_{21}f).
   \end{eqnarray*}   
   Hence we have on $\DD$:\quad
   $[H_{l_1,l_2,l_3},a(f)^*]=l_2a(H_{11}f)^*H_{l_1,l_2-1,l_3}
     +2l_3H_{l_1,l_2,l_3-1}a(\7H_{21}f)$.

   Taking into account
   $$\begin{array}{l}
     (-1)^la(p_1)^*\cdots a(p_l)^*a(p) = a(p)a(p_1)^*\cdots a(p_l)^* 
       +\sum_{j=1}^l(-1)^j\delta(p-p_j)a(p_1)^*\cdots
       \WT{a(p_j)^*}\cdots a(p_l)^*, \\
     a(q_l)^*\cdots a(q_1)^* a(p) = (-1)^la(p)a(q_l)^*\cdots a(q_1)^*
       +\sum_{j=1}^l(-1)^{j-1}\delta(p-q_j)
       a(q_l)^*\cdots\WT{a(q_j)^*}\cdots a(q_1)^*
   \end{array}$$
   and
   $$\begin{array}{rclll}
      \int\4{g(p)}H_{11}(p,q)a(q)\,dp\,dq &=& a({H_{11}}^*g) &=&
        -a(\4{H_{22}}g), \\
      \int a(p)^*H_{12}(p,q)\4{g(q)}\,dp\,dq &=& a(H_{12}\7g)^*, && \\
      \int\4{g(p)}H_{12}(p,q)a(q)^*\,dp\,dq &=& a(\7{H_{12}}^*g)^* &=&
         -a(H_{12}\7g)^*, 
   \end{array}$$
   we find in a similar way:\quad
   $[H_{l_1,l_2,l_3},a(g)]=2l_1a(H_{12}\7g)^*H_{l_1-1,l_2,l_3}+
     l_2H_{l_1,l_2-1,l_3} a(\4{H_{22}}g)$.

   Combination of these commutation relations with
   (\ref{WICK4})--(\ref{AS}) now yields 
   \begin{eqnarray*}
      [\EH,a(f)^*] &=& \sum_{l=0}^\infty 2^{-l}
         \sum_{l_1+l_2+l_3=l}
         \frac{2^{l_2}}{l_1!l_2!l_3!}[H_{l_1,l_2,l_3},a(f)^*] \\
      &=& a(H_{11}f)^* \sum_{l=1}^\infty 2^{-(l-1)} \sum_{l_1+l_2+l_3=l}
         \frac{2^{l_2-1}}{l_1!(l_2-1)!l_3!} H_{l_1,l_2-1,l_3} \\
      && {}+\sum_{l=1}^\infty 2^{-(l-1)} \sum_{l_1+l_2+l_3=l}
         \frac{2^{l_2}}{l_1!l_2!(l_3-1)!}H_{l_1,l_2,l_3-1} a(\7H_{21}f) \\
      &=& a(H_{11}f)^*\EH+\EH a(\7H_{21}f), \\
      && {}\\{}
      [\EH,a(g)] &=& \sum_{l=0}^\infty 2^{-l}
         \sum_{l_1+l_2+l_3=l}
         \frac{2^{l_2}}{l_1!l_2!l_3!}[H_{l_1,l_2,l_3},a(g)] \\
      &=& a(H_{12}\7g)^* \sum_{l=1}^\infty 2^{-(l-1)}
         \sum_{l_1+l_2+l_3=l}
         \frac{2^{l_2}}{(l_1-1)!l_2!l_3!}H_{l_1-1,l_2,l_3} \\
      && {}+\sum_{l=1}^\infty 2^{-(l-1)} \sum_{l_1+l_2+l_3=l}
         \frac{2^{l_2-1}}{l_1!(l_2-1)!l_3!} H_{l_1,l_2-1,l_3}
         a(\4{H_{22}}g) \\
      &=& a(H_{12}\7g)^*\EH+\EH a(\4{H_{22}}g).
   \end{eqnarray*}
\end{Proof}
We next introduce the {\em\ASS e\/}~\cite{R78} $\LV$ of $V$ as a
special solution of (\ref{INTER1})--(\ref{INTER3})\footnote{The \OP s
  $H_{12}$ described below may equivalently be characterized as
  follows. According to Lemma~\ref{lem:BP}, each \aS \hS\oP $T$ from
  $\KK_1$ to $\KK_2$ induces a \bP $P_T$. Then $V^*P_TV=W^*P_1W$ (see
  Lemma~\ref{lem:W}) holds \iFF $-T^*=H_{12}$ for some $H$ as in
  Lemma~\ref{lem:L}.}. Since the ranges of the semi--\fR\OP s $V_{11}$
and ${V_{11}}^*$ are closed, the bounded bijection
$V_{11}|_{\RAN{V_{11}}^*}$ from $\RAN{V_{11}}^*=(\ker V_{11})^\bot$
onto $\RAN V_{11}$ has a bounded inverse. Let
${V_{11}}^{-1}\in\B{\KK_1}$ equal this inverse on $\RAN{V_{11}}$ and
equal zero on $\ker {V_{11}}^*$. We then have
\begin{equation} \label{INVREL}
  \begin{array}{ll}
    \RAN{V_{11}}^{-1}=\RAN{V_{11}}^*, & \ker{V_{11}}^{-1}=\ker{V_{11}}^*,\\
    V_{11}{V_{11}}^{-1}=P_{\rAN V_{11}}, & {V_{11}}^{-1}V_{11}=
      P_{{\rAN V_{11}}^*}
  \end{array}
\end{equation}
where $P_\HH$ denotes the orthogonal projection onto a closed subspace
$\HH\subset\KK$. Of course, the analogous relations hold true for 
${V_{22}}^{-1}=\7{V_{11}}^{-1}\7$. As a generalization of \RUI'
definition in~\cite{R78}, we now set
\begin{equation} \label{LL}
   \begin{array}{l}
      \LV_{12}:=V_{12}{V_{22}}^{-1}-{V_{11}}^{-1*}{V_{21}}^*
        P_{{\ker V_{22}}^*} \\
      \LV_{11}:={V_{11}}^{-1*}-P_1-P_{{\ker V_{11}}^*}
        V_{12}{V_{22}}^{-1}V_{21} \\
      \LV_{22}:=P_2-{V_{22}}^{-1}+{V_{12}}^*{V_{11}}^{-1*}{V_{21}}^*
        P_{{\ker V_{22}}^*} \\ 
      \LV_{21}:=({V_{22}}^{-1}-{V_{12}}^*{V_{11}}^{-1*}{V_{21}}^*
        P_{{\ker V_{22}}^*})V_{21}.
   \end{array}
\end{equation}
\begin{Lem}\label{lem:L}
  The \aS solutions $H$ of (\ref{INTER1})--(\ref{INTER3}) with $H_{12}$ \hS are
  precisely the \OP s of the form
  $$H=\LV+\MAT{-h_{12}V_{21}}{h_{12}}{{V_{12}}^*h_{12}V_{21}}
    {-{V_{12}}^*h_{12}}$$
  where $h_{12}$ is an \aS\hS\oP from $\KK_2$ to $\KK_1$ with
  $$(\ker h_{12})^\bot\subset\ker{V_{22}}^*\ominus V_{21}(\ker V_{11}),
    \qquad\RAN h_{12}\subset\ker{V_{11}}^*\ominus V_{12}(\ker V_{22}).$$
  The space spanned by such \OP s $h_{12}$ has
  \dIM$(m^2-m)/2,\ \ m:=\IND {V_{11}}^*=\frac{1}{2}\IND V^*$.
\end{Lem}
\begin{Proofspc}
  We first note that a Wick ordered expression 
  $a(f)^*\EH+\EH a(g)$ vanishes \iFF $f=g=0$. In fact, application to
  the vacuum gives $a(f)^*\exp(\frac{1}{2}H_{12}a^*a^*)\Omega_{P_1}$
  which equals zero \iFF$f=0$ (to see this, look for instance at the
  one--particle component).  Similarly, $(\EH
  a(g))\,a(g)^*\Omega_{P_1}=\|g\|^2
  \exp(\frac{1}{2}H_{12}a^*a^*)\Omega_{P_1}$ vanishes \iFF$g=0$.

  Hence we get all solutions of (\ref{INTER1})--(\ref{INTER3}) if we
  write these equations in Wick ordered form and then compare
  term by term. We have by Lemma~\ref{lem:REL}
  \begin{eqnarray*}
    \EH a(f)^* &=& a\big((P_1+H_{11})f\big)^*\EH+\EH a(\7H_{21}f), \\
    a_V(f)^*\EH &=& a\big((V_{11}-H_{12}V_{21})f\big)^*\EH+\EH
      a\big(\7(P_2-H_{22})V_{21}f\big), \\
    a_V(g)\EH &=& a\big((V_{12}-H_{12}V_{22})\7g\big)^*\EH+\EH
      a\big((P_1-\4{H_{22}})V_{11}g\big).
  \end{eqnarray*}
  Thus (\ref{INTER1}) is equivalent to
  \begin{eqnarray}
    (P_1+H_{11}-V_{11}+H_{12}V_{21})f &=& 0, \label{LI}\\
    (H_{21}+(H_{22}-P_2)V_{21})f &=& 0,\quad f\in\KK_1, \label{LII}
  \end{eqnarray}
  (\ref{INTER2}) is equivalent to
  \begin{eqnarray}
    (H_{12}V_{22}-V_{12})\7g &=& 0, \label{LIII}\\
    (P_1+(\4{H_{22}}-P_1)V_{11})g &=& 0,\quad 
      g\in\RAN{V_{11}}^*, \label{LIV}
  \end{eqnarray}
  whereas (\ref{INTER3}) is equivalent to 
  \begin{equation} \label{LV}
    \begin{array}{rcl} 
      (P_1+H_{11})h &=& 0, \\
      H_{21}h &=& 0,\quad h\in\ker V_{11}.
    \end{array}
  \end{equation}

  Next we show that each \aS\hS\oP $H_{12}$ fulfilling (\ref{LIII})
  and $V_{21}(\ker V_{11})\subset\ker H_{12}$ gives rise to a unique
  solution $H$ of (\ref{AS}) and (\ref{INTER1})--(\ref{INTER3}). Given
  $H_{12}$, $H_{11}$ is fixed by (\ref{LI}) which in turn yields
  $H_{22}=-\4{{H_{11}}^*}$ by (\ref{AS}). $H_{21}$ is then determined
  by (\ref{LII}) which proves uniqueness of $H$. Explicitly, we have
  \begin{equation} \label{HCOMP}
    H_{11}=V_{11}-P_1-H_{12}V_{21},\qquad
    H_{22}=P_2-{V_{22}}^*-{V_{12}}^*H_{12},\qquad
    H_{21}=({V_{22}}^*+{V_{12}}^*H_{12})V_{21}.
  \end{equation}
  To see that $H$ indeed is a solution of (\ref{AS}) and
  (\ref{INTER1})--(\ref{INTER3}), we have to check antisymmetry of
  $H_{21}$, (\ref{LIV}) and (\ref{LV}) (the rest is clear by
  construction). By antisymmetry of $H_{12}$ and (\ref{REL4}), we have
  $$H_{21}+\4{{H_{21}}^*}=({V_{22}}^*+{V_{12}}^*H_{12})V_{21}+
  {V_{12}}^*(V_{11}-H_{12}V_{21})=0,$$ so $H_{21}$ is \AS. By
  (\ref{REL1}) and (\ref{LIII}), we have for $g\in\RAN{V_{11}}^*$
  $$(P_1+(\4{H_{22}}-P_1)V_{11})g=
    (P_1-({V_{11}}^*+{V_{21}}^*\4{H_{12}})V_{11})g=
    {V_{21}}^*\7(V_{12}-H_{12}V_{22})\7g=0,$$ 
  so (\ref{LIV}) holds. Using (\ref{REL4}), we find for $h\in\ker V_{11}$
  $$(P_1+H_{11})h=-H_{12}V_{21}h,\qquad H_{21}h={V_{12}}^*H_{12}V_{21}h.$$
  Thus (\ref{LV}) is equivalent to $V_{21}(\ker V_{11})\subset\ker
  H_{12}$ which holds by assumption, so $H$ solves (\ref{AS}) and
  (\ref{INTER1})--(\ref{INTER3}).
  
  Finally, we have to characterize the \aS\hS\OP s $H_{12}$ fulfilling
  (\ref{LIII}) and $V_{21}(\ker V_{11})\subset\ker H_{12}$. Note that
  $\LV_{12}$ is \hS since $V_{12}$ is, that $\LV_{12}$ solves
  (\ref{LIII}) and that 
  $$\LV_{12}V_{21}h=-{V_{11}}^{-1*}{V_{21}}^*V_{21}h=-{V_{11}}^{-1*}h=0,
    \qquad h\in\ker V_{11}=\ker{V_{11}}^{-1*}$$
  by (\ref{KER}), (\ref{INVREL}) and (\ref{REL1}). $\LV_{12}$ is also
  \AS: 
  \begin{eqnarray*}
    \LV_{12}+\4{{\LV_{12}}^*} &=& V_{12}{V_{22}}^{-1}-{V_{11}}^{-1*} 
      {V_{21}}^*P_{{\ker V_{22}}^*}+{V_{11}}^{-1*}{V_{21}}^*-
      P_{{\ker V_{11}}^*}V_{12}{V_{22}}^{-1} \\
    &=& P_{\RAN V_{11}}V_{12}{V_{22}}^{-1}+{V_{11}}^{-1*}{V_{21}}^*
      P_{\RAN V_{22}} \\
    &=& {V_{11}}^{-1*}({V_{11}}^*V_{12}+{V_{21}}^*V_{22})
      {V_{22}}^{-1}\\
    &=& 0
  \end{eqnarray*}
  by (\ref{INVREL}) and (\ref{REL3}). Hence $\LV_{12}$ has all the
  desired properties, and one readily checks (using $P_{{\ker
  V_{22}}^*}=P_2-V_{22}{V_{22}}^{-1}$) that the \cOR solution
  of (\ref{AS}) and (\ref{INTER1})--(\ref{INTER3}) is given by (\ref{LL}).

  To find the general form of $H_{12}$, we make the ansatz
  $H_{12}=\LV_{12}+h_{12}$. Then $H_{12}$ is an \aS\hS\oP\iFF $h_{12}$
  is. It fulfills (\ref{LIII}) \iFF $h_{12}V_{22}=0$, and we have
  $V_{21}(\ker V_{11})\subset\ker H_{12}$ \iFF $V_{21}(\ker
  V_{11})\subset\ker h_{12}$. The last two conditions are equivalent
  to $(\ker h_{12})^\bot\subset(\RAN V_{22}\2V_{21}(\ker V_{11}))^\bot
  =\ker{V_{22}}^*\ominus V_{21}(\ker V_{11})$, and we then have by
  antisymmetry $\RAN h_{12}=\7(\RAN{h_{12}}^*)\subset\7(\ker
  h_{12})^\bot \subset\ker{V_{11}}^*\ominus V_{12}(\ker V_{22})$. As a
  result, the admissible components $H_{12}$ (as well as the remaining
  components (\ref{HCOMP})) have the form stated in the lemma. By
  (\ref{DIMKER}), we may regard the $h_{12}$ as \aS\OP s from one
  $m$--\DIM al space to another, thus there are $(m^2-m)/2$ linearly
  independent ones.
\end{Proofspc}
For $H_{12}=\LV_{12}+h_{12}$ as above, we have $\|H_{12}\|_2^2=
\|\LV_{12}\|_2^2+\|h_{12}\|_2^2$. Hence the \ASS e $\LV$ is the \aS
solution of (\ref{INTER1})--(\ref{INTER3}) with minimal \hS norm
$\|\LV_{12}\|_2^{}$. In the rest of Section~\ref{sec:CON}, we shall
work exclusively with $\LV$. But we emphasize that {\em all the
  results in Sections~\ref{sec:NORMFORM} and \ref{sec:DEC} hold as
  well if $\LV$ is replaced everywhere by one of the \OP s $H$
  described in Lemma~\ref{lem:L}}. The only point where the choice of
the \ASS e (of $W$ below, not of $V$) is distinguished is in
Section~\ref{sec:DEC} ($\LW_{12}=0$, see Lemma~\ref{lem:W}). We remark
further that (\ref{LL}) reduces to the definition given by
\RUI~\cite{R78} (i.e.~$\LV_{12}=V_{12}{V_{22}}^{-1}$) whenever
$\OM_{P_1}\0\R{V}$ is
pure (cf.\ Lemma~\ref{lem:PURE}~a)). 
\subsection{Normal Form of Implementers} \label{sec:NORMFORM}
As we have seen in Section~\ref{sec:BH}, $\EV$ is (the \qF of) a
densely defined \oP with intertwining properties
(\ref{INTER1})--(\ref{INTER3}). To construct an \ISM ic \IMP er for
$\R{V}$, let $$L_V:=\dim\ker V_{11}<\infty$$ (see~(\ref{LVFIN})) and
choose an \oNB $\{e_1,\ldots,e_{L_V}\}$ for $\ker V_{11}$. For
$r=1,\ldots,L_V$, set\footnote{It is also possible to incorporate the
  factors $\PME$ into $\EV$ as is done in~\cite{R78,CB}, but the
  choice (\ref{AV}) simplifies the combinatorics.}
\begin{equation} \label{AV}
  \begin{array}{l}
    A_r:=a(e_r)\PME,\\
    A_{V,r}:=a_V(e_r)\PME=a(V_{12}\7e_r)^*\PME
  \end{array}
\end{equation}
where $\PME$ is the self--adjoint \UNI y \IMP er for $\8_{-1}$ with
$\PME\Omega_{P_1}=\Omega_{P_1}$ (cf.\ the proof of
Lemma~\ref{lem:MIN}). Then the ${A_r}^{(*)},\ {A_{V,r}}^{(*)}$
respectively fulfill the CAR\@. Let $\PLV$ denote the index set
consisting of pairs $(\6,s)$ with $s\in\{0,\ldots,L_V\}$ and $\6$ a
permutation of order $L_V$ satisfying $\6(1)<\ldots<\6(s)$ and
$\6(s+1)<\ldots<\6(L_V)$. $\PLV$ is canonically isomorphic to the
power set $\PS$ of $\{1,\ldots,L_V\}$ through identification of $(\6,s)$ with
$\{\6(1),\ldots,\6(s)\}$, hence its cardinality is $2^{L_V}$.  We now
define the following \oP on $\DD$
\begin{eqnarray}
  \PNV &:=& \Big(\det\big(P_1+\LV_{12}{\LV_{12}}^*\big)\Big)^{-1/4}
    \cdot\label{PNV} \\
  && \cdot\sum_{(\6,s)\in\PLV}(-1)^s\SIGS\ A_{V,\6(1)}\cdots 
  A_{V,\6(s)} \EV A_{\6(s+1)}\cdots A_{\6(L_V)}
\end{eqnarray}
with range contained in the space of $C^\infty$--vectors for the
number \OP.
\begin{Lem}\label{lem:PSINULL}
  $\PNV$ has a continuous extension to an \ISM y (denoted by the same
  symbol) on $\FA{\KK_1}$ with
  $$\PNV\pi_{P_1}(A)=\pi_{P_1}(\R{V}(A))\PNV,\quad A\in\CK.$$
\end{Lem}
\begin{Proofspc}
  We first show 
  \begin{equation} \label{INTER4}
    \PNV a(f)^{(*)}=a_V(f)^{(*)}\PNV,\quad f\in\KK_1
  \end{equation}
  on $\DD$. To this end, let us introduce the analog of \RUI' \oP
  $\hat\7(V)$~\cite{R78}
  \begin{equation}
    \label{PV}
    \PV:=\Big(\det\big(P_1+\LV_{12}{\LV_{12}}^*\big)\Big)^{-1/4}\EV.
  \end{equation}
  For $f\in(\ker{V_{11}})^\bot$, (\ref{INTER4}) follows from (\ref{INTER1}),
  (\ref{INTER2}) together with
  $[a(f)^{(*)},A_r]=[a_V(f)^{(*)},A_{V,r}]=0$. To prove (\ref{INTER4})
  for $f=e_r,\ r=1,\ldots,L_V$, we make use of
  \begin{eqnarray}
    && [a(e_r),A_s]=[a_V(e_r),A_{V,s}]=0, \label{RELA1}\\{}
    && [a(e_r)^*,A_s]=[a_V(e_r)^*,A_{V,s}]=\delta_{rs}\PME, \label{RELA2}\\
    && \{\PME,A_s\}=\{\PME,A_{V,s}\}=[\PME,\PV]=0. \label{RELA3}
  \end{eqnarray}
  Note further that for fixed $r$, the bijection $\{\MM\in
  \PS\,|\,r\in\MM\}\to \{\MM'\in \PS\,|\,r\notin\MM'\},\ 
  \MM\mapsto\MM\setminus\{r\}$ induces a bijection
  $(\6,s)\mapsto(\6',s')$ from
  $\{(\6,s)\in\PLV\,|\,r\in\{\6(1),\ldots,\6(s)\}\}$ onto
  $\{(\6',s')\in\PLV\,|\,r\notin\{\6'(1),\ldots,\6'(s')\}\}$ 
  with~\cite{CB}
  \begin{equation} \label{RELBIJ}
    s=s'+1,\quad(-1)^s\SIGS=(-1)^r\SIGS',\quad\6^{-1}(r)+\6'^{-1}(r)=r+s.
  \end{equation}
  We now have by virtue of the CAR, (\ref{RELA1}), (\ref{RELA3}) and
  (\ref{RELBIJ}) on $\DD$
  \begin{eqnarray*}
    \PNV a(e_r) &=& A_{V,r}\sum_{(\6,s)\in\PLV\atop
      r\in\{\6(1),\ldots,\6(s)\}}(-1)^{s+\6^{-1}(r)-1}\SIGS\ 
      A_{V,\6(1)}\cdots\WT{A_{V,r}}\cdots A_{V,\6(s)} \cdot\\
    &&{}\cdot\PV A_{\6(s+1)}\cdots a(e_r)\PME^2\cdots A_{\6(L_V)} \\
    &=& a_V(e_r)\PME^2\sum_{(\6',s')\in\PLV\atop
      r\notin\{\6'(1),\ldots,\6'(s')\}}(-1)^{s'}\SIGS'\ 
      A_{V,\6'(1)}\cdots A_{V,\6'(s')}\PV
      A_{\6'(s'+1)}\cdots A_{\6'(L_V)} \\
    &=& a_V(e_r)\PNV.
  \end{eqnarray*}
  As a consequence of (\ref{INTER1}) and (\ref{INTER3}), we have $\PV
  a(e_r)^*=a_V(e_r)^*\PV=0$. This yields in \CON ion with
  (\ref{RELA2}), (\ref{RELA3}) and (\ref{RELBIJ})
  \begin{eqnarray*}
    \lefteqn{a_V(e_r)^*\PNV=} \\
    &=& \PME\sum_{(\6,s)\in\PLV\atop r\in\{\6(1),\ldots,\6(s)\}}
      (-1)^{s+\6^{-1}(r)-1}\SIGS\ A_{V,\6(1)}\cdots\WT{A_{V,r}}
      \cdots A_{V,\6(s)}\PV A_{\6(s+1)}\cdots A_{\6(L_V)} \\
    &=& \PME\sum_{(\6',s')\in\PLV\atop
      r\notin\{\6'(1),\ldots,\6'(s')\}}(-1)^{s'+\6'^{-1}(r)}\SIGS'\ 
      A_{V,\6'(1)}\cdots A_{V,\6'(s')}\PV
      A_{\6'(s'+1)}\cdots\WT{A_r}\cdots A_{\6'(L_V)} \\
    &=& \PNV a(e_r)^*,
  \end{eqnarray*}
  so (\ref{INTER4}) holds. 

  Since the ${A_{V,r}}^{(*)}$ fulfill the CAR and ${A_{V,r}}^*\PV=0$,
  \RUI' result~(\ref{OMNORM}) implies
  \begin{eqnarray*}
    \|\PNV\Omega_{P_1}\|^2 &=& \|A_{V,1}\cdots
      A_{V,L_V}\PV\Omega_{P_1}\|^2\\ 
    &=& \<{\PV\Omega_{P_1},{A_{V,L_V}}^*A_{V,L_V}\cdots{A_{V,1}}^*A_{V,1}
      \PV\Omega_{P_1}} \\
    &=& \|\PV\Omega_{P_1}\|^2 \\
    &=& 1.
  \end{eqnarray*}
  Since the ${a_V(f)}^{(*)}$ also fulfill the CAR and since
  ${a_V(f)}\PNV\Omega_{P_1}=0$ by (\ref{INTER4}), we obtain for
  $g_1,\ldots,g_m,h_1,\ldots,h_n\in\KK_1$
  \begin{eqnarray*}
    \lefteqn{\<{\PNV a(g_1)^*\cdots a(g_m)^*\Omega_{P_1},\PNV a(h_1)^*
      \cdots a(h_n)^*\Omega_{P_1}}=} \\
    &=& \<{\PNV\Omega_{P_1},a_V(g_m)\cdots 
      a_V(g_1)a_V(h_1)^*\cdots a_V(h_n)^*\PNV\Omega_{P_1}} \\
    &=& \<{a(g_1)^*\cdots a(g_m)^*\Omega_{P_1},
      a(h_1)^*\cdots a(h_n)^*\Omega_{P_1}}.
  \end{eqnarray*}
  Hence $\PNV$ is \ISM ic on $\DD$ and has a continuous extension to
  an \ISM y which satisfies (\ref{INTER4}) on $\FA{\KK_1}$. But this
  implies $\PNV\pi_{P_1}(A)=\pi_{P_1}(\R{V}(A))\PNV$ for $A\in\CK$.
\end{Proofspc}
We proceed to construct a complete set of \IMP ers with the help of
$\PNV$. In view of the remark above Lemma~\ref{lem:DEC}, we have to
look for partial \ISM ies in $\pi_{P_1}(\R{V}(\CK))'$ which contain
$\RAN\PNV$ in their initial spaces. Since $\KK$ is infinite \DIM al,
we have~\cite{A70} $\pi_{P_1}(\R{V}(\CK))'=\psi(\ker V^*)''$ with
$$\psi(k):=\pi_{P_1}(B(k))\PME.$$ 
\begin{Lem}
  Let $k\in\ker V^*$. Then $\psi(k)$ is a partial \ISM y with
  $\RAN\PNV\subset(\ker\psi(k))^\bot$ \iFF a) or b) below holds.\\
  a) $\|k\|=\sqrt{2},\ |\<{k,\7k}|=2$. In this case, $\psi(k)$ is
    \UNI y.\\ 
  b) $\|k\|=1,\ k\in\RAN(P_1-{\LV_{12}}^*)$. In this case,
    $\FA{\KK_1}=(\ker\psi(k))^\bot\2\RAN\psi(k)$.
\end{Lem}
\begin{Proofspc}
  Let $k\in\KK$. $\psi(k)^*\psi(k)$ and $\psi(k)\psi(k)^*$ are
  projections \iFF one of the following holds:
  $$1)\ \ k=0,\qquad 2)\ \ \|k\|^2=|\<{k,\7k}|=2,\qquad 3)\ \ \|k\|=1,\
    \<{k,\7k}=0.$$  
  In the second case we have $|\<{k,\7k}|=\|k\|\cdot\|\7k\|$, hence
  there exists $z\in U(1)$ with $\7k=zk$. This implies
  $B(k)^*B(k)=B(\7k)B(k)=\frac{z}{2}\{B(k),B(k)\}
    =\frac{z}{2}\<{\7k,k}\1=\1$ 
  and $B(k)B(k)^*=\|k\|^2\1-B(k)^*B(k)=\1$, thus $\psi(k)$ is \UNI y.

  In the third case we have $\psi(k)^*\psi(k)=\1-\psi(k)\psi(k)^*$,
  hence initial and final space of $\psi(k)$ are orthogonal to each
  other and sum up to $\FA{\KK_1}$. The requirement
  $\RAN\PNV\subset(\ker\psi(k))^\bot=\ker\psi(k)^*$ holds for
  $k\in\ker V^*$ \iFF $\psi(k)^*\PNV\Omega_{P_1}=0$. This follows from
  $\psi(k)^*\in\pi_{P_1}(\R{V}(\CK))'$ and the fact that vectors of
  the form $\PNV\pi_{P_1}(A)\Omega_{P_1}= \pi_{P_1}(\R{V}(A))
  \PNV\Omega_{P_1},\ A\in\CK$, are dense in $\RAN\PNV$.  By
  $\{\psi(k)^*,A_{V,r}\}=0$, we further have (cf.\ the proof of
  Lemma~\ref{lem:PSINULL}) $\|\psi(k)^*\PNV\Omega_{P_1}\|=\|\psi(k)^*
  \PV\Omega_{P_1}\|$. By Lemma~\ref{lem:REL} and (\ref{RELA3}),
  $$\psi(k)^*\PV\Omega_{P_1}=-(a(P_1k)+a(P_1\7k)^*)\PV\Omega_{P_1}=
    -a((P_1-\LV_{12})\7k)^*\PV\Omega_{P_1}$$ which vanishes \iFF
  $k\in\ker(P_1-\LV_{12})\7=\RAN(P_1-{\LV_{12}}^*)$ (cf.\ 
  (\ref{RANKER}) below). But for such $k$, 
  $\<{k,\7k}=0$ automatically holds (see Section~\ref{sec:DEC}), so we
  conclude that partial \ISM ies $\psi(k)$ of type~3) with $k\in\ker
  V^*$ and $\RAN\PNV\subset(\ker\psi(k))^\bot$ are completely
  characterized by condition b).
\end{Proofspc}
For our purposes, the partial \ISM ies described in part b) of the
lemma are the important ones. Let $\{k_1,\ldots,k_m\}$ be an \oNB for
$\ker V^*\cap\RAN(P_1-{\LV_{12}}^*)$. For $\3=(\3_1,\ldots,\3_r)\in
I_m$ (cf.\ (\ref{I}) and (\ref{DEF})), set 
\begin{equation} \label{PSIB}
  \begin{array}{l}
    \psi_\3:=\psi(k_{\3_1})\cdots\psi(k_{\3_r}), \\
    \Psi_\3(V):=\psi_\3\PNV.
  \end{array}
\end{equation}
Since the $(\psi_j^{(*)})_{j=1,\ldots,m}$ fulfill the CAR
\begin{equation} \label{PSICAR}
  \{\psi_j,\psi_l\}=\{\psi_j^*,\psi_l^*\}=0,\qquad
  \{\psi_j,\psi_l^*\}=\delta_{jl}\1,
\end{equation}
the $\psi_\3$ are partial \ISM ies in $\pi_{P_1}(\R{V}(\CK))'$ with
$\RAN\PNV\subset(\ker\psi_\3)^\bot$.
\begin{Th}\label{th:IMPISM}
  $\ker V^*\cap\RAN(P_1-{\LV_{12}}^*)$ has \dIM $m=\frac{1}{2}\IND
  V^*$, and the $d_V=2^m$ \ISM ies $(\Psi_\3(V))_{\3\in I_m}$ \iMP
  $\R{V}$ in $\pi_{P_1}$ in the sense of Definition~\ref{def:IMP}.
\end{Th}
\begin{Proofspc}
  As a consequence of (\ref{PSICAR}) and $\psi_j^*\PNV=\PNV^*\psi_j=0$,
  the first equation in (\ref{CUNTZ}) holds:
  \begin{equation} \label{CUNTZ0}
    \Psi_\3(V)^*\Psi_\5(V)=\PNV^*\psi_{\3_r}^*\cdots\psi_{\3_1}^*
    \psi_{\5_1}\cdots\psi_{\5_s}\PNV=\delta_{\3\5}\1
  \end{equation}
  for $\3=(\3_1,\ldots,\3_r),\5=(\5_1,\ldots,\5_s)\in I_m$. Clearly,
  the $\Psi_\3(V)$ have the intertwining property~(\ref{INTER})
  \begin{equation} \label{CUNTZ1}
    \Psi_\3(V)\pi_{P_1}(A)=\pi_{P_1}(\R{V}(A))\Psi_\3(V),\quad A\in\CK
  \end{equation}
  since $\psi_\3\in\pi_{P_1}(\R{V}(\CK))'$. We postpone the proofs of
  the completeness relation
  \begin{equation} \label{CUNTZ2}
    \sum_{\3\in I_m}\Psi_\3(V)\Psi_\3(V)^*=\1
  \end{equation}
  and of $m=\frac{1}{2}\IND V^*$ to Section~\ref{sec:DEC}. Since
  (\ref{CUNTZ1}) and (\ref{CUNTZ2}) imply (\ref{IMPEND}), the theorem
  will then be proven.
\end{Proofspc}
By (\ref{PSICAR}) and by $\psi_j^*\PNV=0,\ j=1,\ldots,m$, the $\psi_j$
may be regarded as creation \OP s relative to the vacuum $\PNV$. The
\hSP spanned by the $\Psi_\3(V)$ is therefore canonically isomorphic
to the \aS Fock space over $\ker V^*\cap\RAN(P_1-{\LV_{12}}^*)$.
\subsection{Decomposition of \bOG Operators and Proof of Completeness}
\label{sec:DEC}
Let us first remark that $m:=\dim(\ker V^*\cap\RAN(P_1-{\LV_{12}}^*))
=\frac{1}{2}\IND V^*$ implies completeness~(\ref{CUNTZ2}) in the case
of finite index since the \rEP $\pi_{P_1}\0\R{V}$ splits into $d_V$
\IRR s by Theorem~\ref{th:DEC} and since the ranges of the \ISM ies
$\Psi_\3(V)$ are mutually orthogonal, \iRR subspaces for
$\pi_{P_1}\0\R{V}$.  However, we shall give a different proof of
completeness which also works in the case of infinite index. The goal
is a (to a certain extent {\em canonical\/}) product \dEC $V=UW$ where
$U\in\INK{0}{P_1}$ is \UNI y and $W\in\INK{\iND V^*}{P_1}$ induces a
pure and gauge \iNV state $\OM_{P_1}\0\R{W}$. $U$ and $W$ will be
chosen such that $\LU_{12}=\LV_{12}$ and $\LW_{12}=0$, and
(\ref{CUNTZ2}) will follow from completeness of \IMP ers for $\R{W}$
which in turn is a consequence of Lemma~\ref{lem:DEC}.

We start with the proof of  $m=\frac{1}{2}\IND V^*$.
\begin{Lem}\label{lem:BP}
  Let $T$ be an \aS\hS\oP from $\KK_1$ to $\KK_2$. Then
  $\mAT{P_1}{\4{T}}{T}{P_2}$ is a bijection on $\KK$, and
  $\KK=\RAN(P_1+T)\2\RAN(P_2+\4{T})$. If we set
  \begin{eqnarray*}
    P &:=& (P_1+T)(P_1+T^*T)^{-1}(P_1+T^*),\\
    U_T &:=& (P_1+T)(P_1+T^*T)^{-1/2}+(P_2+\4{T})(P_2+TT^*)^{-1/2},
  \end{eqnarray*}
  then $P$ is a \bP with $\RAN P=\RAN(P_1+T)$ and $P_2P\in\JK{2}$, and
  $U_T\in\INK{0}{P_1}$ is \UNI y with $U_T^*PU_T=P_1$.
\end{Lem}
\begin{Proofspc}
  Let $k\in\ker(\1+T+\4{T})$. Then $P_1k=-\4{T}P_2k$ and
  $P_2k=-TP_1k$, hence $(P_1+T^*T)P_1k=0$ by antisymmetry~(\ref{AS}).
  But $P_1+T^*T$ is a bijection on $\KK_1$, so $k=0$ and $\1+T+\4{T}$
  is injective. Since $\1+T+\4{T}$ is \fR with vanishing index by
  compactness of $T$, it is also surjective.

  Let $f_j\in\KK_j,\ j=1,2$. Then $\<{(P_1+T)f_1,(P_2+\4{T})f_2}
  =\<{f_1,\4{T}f_2}+\<{Tf_1,f_2}=0$ by antisymmetry which proves
  $\KK=\RAN(P_1+T)\2\RAN(P_2+\4{T})$. It is not hard to see that $P$
  is the projection onto $\RAN(P_1+T)$ and therefore a \BP. The \UNI y
  $U_T$ results from polar \dEC of
  $\1+T+\4{T}=U_T$\mbox{$|\1+T+\4{T}|$} (by the way, $U_T$ coincides
  with Araki's canonical choice of a \bOG\oP that transforms $P$ into
  $P_1$~\cite{A87}). $P_2P=T(P_1+T^*T)^{-1}(P_1+T^*)$ and
  $(U_T)_{21}=T(P_1+T^*T)^{-1/2}$ are \hS since $T$ is, and
  $PU_T=(P_1+T)(P_1+T^*T)^{-1/2}=U_TP_1$.
\end{Proofspc}
Application of Lemma~\ref{lem:BP} to $T=\4{\LV_{12}}=-{\LV_{12}}^*$ yields
the \bP $P_V$ with $\RAN P_V=\RAN(P_1-{\LV_{12}}^*)$.
\begin{Lem}\label{lem:PV}
  $P_V$ commutes with $VV^*$. As a consequence, $\ker V^*=P_V(\ker
  V^*)\2\4{P_V}(\ker V^*)$ and $m=\dim P_V(\ker V^*)=\frac{1}{2}\IND
  V^*$.
\end{Lem}
\begin{Proof}
  $[P_V,VV^*]=0$ is equivalent to $[\4{P_V},VV^*]=0$ or to
  $VV^*\Big(\RAN(P_2+\LV_{12})\Big)\subset\ker(P_1-\LV_{12})$ since
  \begin{equation} \label{RANKER}
    \RAN\4{P_V}=\RAN(P_2+\LV_{12})=\ker P_V=\ker(P_1-\LV_{12}).
  \end{equation} 
  By definition, $\LV_{12}$ fulfills
  $\LV_{12}V_{22}=V_{12}P_{\RAN{V_{22}}^*}$ and $\LV_{12}V_{21}P_{\ker
    V_{11}}=0$. Antisymmetry of $\LV_{12}$ implies
  ${V_{11}}^*\LV_{12}=-P_{\RAN{V_{11}}^*}{V_{21}}^*$ and $P_{\ker
    V_{22}}{V_{12}}^*\LV_{12}=0$. Using these relations, we get
  \begin{eqnarray*}
    (P_1-\LV_{12})VV^*(P_2+\LV_{12}) &=& V_{11}{V_{21}}^*+ V_{12} 
      {V_{22}}^*+(V_{11}{V_{11}}^*+V_{12}{V_{12}}^*)\LV_{12}-\LV_{12}\cdot\\
    && {}\cdot(V_{21}{V_{21}}^*+ V_{22}{V_{22}}^*)
      -\LV_{12}(V_{22}{V_{12}}^*+ V_{21}{V_{11}}^*)\LV_{12}\\
    &=& V_{11}{V_{21}}^*+V_{12}{V_{22}}^*-V_{11}{V_{21}}^*
      +V_{12}{V_{12}}^*\LV_{12}-\LV_{12}V_{21}{V_{21}}^*\\
    && {}-V_{12}{V_{22}}^*-V_{12}P_{\RAN{V_{22}}^*}{V_{12}}^*\LV_{12}+
      \LV_{12}V_{21}P_{\RAN{V_{11}}^*}{V_{21}}^*\\ 
    &=& V_{12}P_{\ker{V_{22}}}{V_{12}}^*\LV_{12}-
      \LV_{12}V_{21}P_{\ker{V_{11}}}{V_{21}}^*\\ 
    &=& 0.
  \end{eqnarray*}
\end{Proof}
Next we present a distinguished choice of $W$ for the product \dEC
$V=UW$. Namely, let $W_{11}$ be the partial \ISM y with $\ker
{W_{11}}^{(*)}=\ker {V_{11}}^{(*)}$ appearing in the polar \dEC of
$V_{11}$:
\begin{equation} \label{W1}
  V_{11}=W_{11}|V_{11}|
\end{equation}
(the idea of using polar \dEC of $V_{11}$ stems from~\cite{CHOB}). Set
\begin{equation} \label{W2}
  W_{21}:=V_{21}P_{\ker{V_{11}}},
\end{equation}
then $W_{21}$ is a partial \ISM y with initial space $\ker V_{11}$ and
final space $V_{21}(\ker V_{11})\subset\ker{V_{22}}^*$
(cf.~(\ref{KER})), and set $W_{12}:=\4{W_{21}},\ W_{22}:=\4{W_{11}}$.
\begin{Lem}\label{lem:W}
  $W$ defined above has the following properties:\\ 
  a) $W\in\IK{P_1}$, $\ker W^*=(\ker{V_{11}}^*\ominus V_{12}(\ker V_{22}))
  \2(\ker{V_{22}}^*\ominus V_{21}(\ker V_{11}))$ and $\IND W=\IND V$;\\ 
  b) $S_W:=W^*P_1W$ is the projection onto $(\ker V_{11})^\bot\2\ker
  V_{22}$, and $\OM_{S_W}$ is pure and gauge \INV;\\ 
  c) $\LW_{12}=0$.
\end{Lem}
\begin{Proofspc}
  a) $W$ is clearly a \bOG\oP with $\ker W^*=(\ker{V_{11}}^*\ominus
  V_{12}(\ker V_{22}))\2(\ker{V_{22}}^*\ominus V_{21}(\ker V_{11}))$.
  Therefore $\IND W^*=\dim\ker W^*=2\IND {V_{11}}^*=\IND V^*$
  (cf.~(\ref{DIMKER})), and $W\in\IK{P_1}$ since $W_{12}$ has finite
  rank.

  b) By a straightforward computation, $S_W$ is the projection onto
  $(\ker V_{11})^\bot\2\ker V_{22}$. $\OM_{S_W}=\OM_{P_1}\0\R{W}$ is
  pure by Lemma~\ref{lem:PURE}~a) and gauge \iNV by $[P_1,S_W]=0$
  (cf.\ Section~\ref{sec:PRE}).
  
  c) $\LW_{12}=0$ follows from ${W_{11}}^{-1}={W_{11}}^*$ and
  $W_{12}{W_{22}}^*=W_{11}{W_{21}}^*=0$.
\end{Proofspc}
It remains to specify the factor $U$ in $V=UW$. $U$ has necessarily
the form $U=VW^*+u$ where $u=\4{u}$ is a partial \ISM y with initial
space $\ker W^*=P_1(\ker W^*)\2P_2(\ker W^*)$ and final space $\ker
V^*=P_V(\ker V^*)\2\4{P_V}(\ker V^*)$. We may choose $u$ such that
$uP_1=P_Vu$ (for example, suitable $uP_1$ is obtained by polar \dEC of
$R_V$ below). Then $P_2P_V\in\JK{2}$ implies that $u_{21}=P_2P_Vu$ and
$U_{21}=V_{21}{W_{11}}^*+u_{21}$ are \HS. We need the following lemma
to exhibit further properties of $U$. Remember that $Q_W=\1-WW^*$
denotes the projection onto $\ker W^*$.
\begin{Lem}\label{lem:RV}
  $R_V:=(P_1-{\LV_{12}}^*)(P_1+V_{11}{V_{21}}^*{\LV_{12}}^*)P_1Q_W$
  maps $P_1(\ker W^*)$ bijectively onto $P_V(\ker V^*)$.
\end{Lem}
\begin{Proof}
  By Lemma~\ref{lem:PV}, $P_V(\ker V^*)=\ker(P_2+{\LV_{12}}^*)\cap\ker
  V^*$. Hence $k\in P_V(\ker V^*)$ \iFF $P_2k=-{\LV_{12}}^*P_1k$,
  ${V_{11}}^*P_1k+{V_{21}}^*P_2k=0$ and
  ${V_{22}}^*P_2k+{V_{12}}^*P_1k=0$. Thus $P_2k$ is determined by
  $P_1k$, and $P_1k$ has to satisfy
  $$P_1k\in\ker\Big({V_{11}}^*-{V_{21}}^*{\LV_{12}}^*\Big)\cap
    \ker\Big({V_{12}}^*-{V_{22}}^*{\LV_{12}}^*\Big).$$  
  $\LV_{12}V_{22}=V_{12}P_{\rAN{V_{22}}^*}$ implies
  $\ker({V_{12}}^*-{V_{22}}^*{\LV_{12}}^*)=
  \RAN(V_{12}-V_{12}P_{\rAN{V_{22}}^*})^\bot=(V_{12}(\ker
  V_{22}))^\bot$. Since 
  \begin{equation}\label{K1DEC}
    \KK_1=P_1(\ker W^*)\2\RAN V_{11}\2V_{12}(\ker V_{22})
  \end{equation}
  (cf.~Lemma~\ref{lem:W}), we may write $P_1k=f+g,\ f\in P_1(\ker
  W^*),\ g\in\RAN V_{11}$. We then have
  ${V_{21}}^*{\LV_{12}}^*g={V_{21}}^*{\LV_{12}}^*V_{11}{V_{11}}^{-1}g=
  -{V_{21}}^*\7\LV_{12}V_{22}\7{V_{11}}^{-1}g=
  -{V_{21}}^*V_{21}P_{\rAN{V_{11}}^*}{V_{11}}^{-1}g
  =({V_{11}}^*-{V_{11}}^{-1})g$ by (\ref{INVREL}) and
  (\ref{REL1}). Hence the condition
  $P_1k\in\ker({V_{11}}^*-{V_{21}}^*{\LV_{12}}^*)$ is equivalent to
  ${V_{11}}^{-1}g={V_{21}}^*{\LV_{12}}^*f$ or to
  $g=V_{11}{V_{21}}^*{\LV_{12}}^*f$ (since $P_{\ker
    V_{11}}{V_{21}}^*{\LV_{12}}^*=0$). As a result, $k\in P_V(\ker V^*)$
  \iFF there exists $f\in P_1(\ker W^*)$ such that
  $P_1k=(P_1+V_{11}{V_{21}}^*{\LV_{12}}^*)f$ and
  $P_2k=-{\LV_{12}}^*P_1k$, i.e.\ $k=R_Vf$. Hence $\RAN R_V=P_V(\ker
  V^*)$. 

  To show that $R_V$ is one--to--one, assume that $f\in
  P_1(\ker W^*)$ and $R_Vf=0$. Then $0=P_1Q_WR_Vf=f$ since
  $P_1Q_WR_V=P_1Q_W$. 
\end{Proof}
\begin{Prop}\label{prop:VUW}
  Let $V\in\IK{P_1}$, and let $W\in\IK{P_1}$ be defined by (\ref{W1})
  and (\ref{W2}) (with properties listed in Lemma~\ref{lem:W}). Then
  there exists $U\in\INK{0}{P_1}$ with $U(P_1(\ker W^*))=P_V(\ker
  V^*)$ and $V=UW$. Such $U$ fulfills $\ker U_{11}=\{0\}$ and
  $\LU_{12}=\LV_{12}$.
\end{Prop}
\begin{Proofspc}
  It remains to prove the last two statements. We have
  $U_{11}=P_{V_{12}(\ker V_{22})}+|{V_{11}}^*|+u_{11}$ by definition
  of $W$, and $\RAN u_{11}=\RAN P_1P_Vu=\RAN P_1R_V\subset P_1(\ker
  W^*)\2\RAN V_{11}$ by $uP_1=P_Vu$ and by Lemma~\ref{lem:RV}. This
  implies $\ker U_{11}\subset P_1(\ker W^*)\2\RAN V_{11}$
  (cf.~(\ref{K1DEC})). Let $f\in P_1(\ker W^*),\ g\in\RAN V_{11}$, and
  assume $0=U_{11}(f+g)=u_{11}f+|{V_{11}}^*|g$. By Lemma~\ref{lem:RV},
  there exists $f'\in P_1(\ker W^*)$ with $uP_1f=R_Vf'$. Then
  $0=P_1Q_WU_{11}(f+g)=P_1Q_WR_Vf'=f'$, hence $f=0=g$. This proves
  $\ker U_{11}=\{0\}$. 

  Let $f\in P_1(\ker W^*)$. Since there exists $f'\in P_1(\ker W^*)$
  with $uP_1f=R_Vf'$, we see that
  $u_{21}f=P_2R_Vf'=\4{\LV_{12}}P_1R_Vf'=\4{\LV_{12}}u_{11}f$ by
  definition of $R_V$. Hence $\LV_{12}U_{22}=(V_{12}{V_{22}}^{-1}-
  {V_{11}}^{-1*}{V_{21}}^*P_{{\ker V_{22}}^*})(P_{V_{21}(\ker
    V_{11})}+|{V_{22}}^*|)+\LV_{12}u_{22} =V_{12}{V_{22}}^{-1}
  |{V_{22}}^*| +u_{12}=U_{12}$ by (\ref{INVREL}) and
  $|{V_{22}}^*|=V_{22}{W_{22}}^*$. But this means $\LU_{12}=\LV_{12}$
  (cf.\ the remark below Lemma~\ref{lem:L}).
\end{Proofspc}
%
The following result has already been obtained, in the case of finite
index, in Lemma~\ref{lem:EQ} ($S_V-S_{V'}$ is automatically \hS for
$V,V'\in\IK{P_1}$). 
\begin{Cor}
  The $\INK{0}{P_1}$--orbits in $\IK{P_1}$ with respect to left
  multiplication are precisely the sets $\INK{2m}{P_1},\
  m\in\NN\cup\{\infty\}$. 
\end{Cor}
\begin{Proofspc}
  Let $V,V'\in\INK{2m}{P_1}$ be given, with \DEC s $V=UW,\ V'=U'W'$ as
  in Proposition~\ref{prop:VUW}. Since $P_1$ leaves $\ker W'^*$ and
  $\ker W^*$ \INV, we may choose a partial \ISM y $u''$ with initial
  space $\ker W'^*$ and final space $\ker W^*$ such that $\4{u''}=u''$
  and $[P_1,u'']=0$. Then $U'':=WW'^*+u''\in\INK{0}{P_1}$ fulfills
  $U''W'=W$. This implies $(UU''U'^*)V'=V$, so $\INK{0}{P_1}$ acts
  transitively on $\INK{2m}{P_1}$.
\end{Proofspc}
{\it End of Proof of Theorem~\ref{th:IMPISM}}. Let $V=UW$ as
in Proposition~\ref{prop:VUW}. We first apply the construction from
Section~\ref{sec:NORMFORM} to $W$ and compute the values of \IMP ers
$\Psi_\3(W)$ on $\Omega_{P_1}$. Since $\ker{W_{11}}=\ker {V_{11}}$ and
$W_{12}|_{\ker V_{22}}=V_{12}|_{\ker V_{22}}$, we may choose
$A_{W,r}=A_{V,r},\ r=1,\ldots,L_V$ (cf.~(\ref{AV})). Then
$\Psi_0(W)\Omega_{P_1}=(-1)^{L_V}A_{V,1}\cdots A_{V,L_V}\Omega_{P_1}$
by $\LW_{12}=0$ (cf.~(\ref{PNV})). According to
Section~\ref{sec:NORMFORM}, we have to
choose an \oNB $\{f_1,\ldots,f_m\}$ for $\ker
W^*\cap\RAN(P_1-{\LW_{12}}^*)=P_1(\ker W^*)$, $m=\frac{1}{2}\IND
W^*=\frac{1}{2}\IND V^*$, to obtain further \IMP ers for
$\R{W}$. Since $U(P_1(\ker W^*))=P_V(\ker V^*)=\ker
V^*\cap\RAN(P_1-{\LV_{12}}^*)$, we may choose the $f_j$ such that
$Uf_j=k_j,\ j=1,\ldots,m$. For a multi--index $\3=(\3_1,\ldots,\3_r)\in
I_m$, we have by definition~(\ref{PSIB})
$$\Psi_\3(W)\Omega_{P_1}=(-1)^{L_V}\psi(f_{\3_1})\cdots\psi(f_{\3_r})
  A_{V,1}\cdots A_{V,L_V}\Omega_{P_1}.$$
Let $A:=a_W(e_{L_V})^*\cdots a_W(e_1)^*\in\pi_{P_1}(\R{W}(\CK))$
(cf.~(\ref{AV})). Remembering
$\psi(f_j)=a(f_j)^*\PME\in\pi_{P_1}(\R{W}(\CK))'$ and neglecting signs, we get
$A\Psi_\3(W)\Omega_{P_1}=\pm a(f_{\3_1})^*\cdots
a(f_{\3_r})^*\Omega_{P_1}=\pm\phi_\3^W$ where $\phi_\3^W\in\FF_\3^W$
is the cyclic vector defined in (\ref{DEF}). Since $\OM_{P_1}\0\R{W}$
is pure, $\FF_\3^W$ is an \iRR subspace for $\pi_{P_1}\0\R{W}$. But by
(\ref{CUNTZ0}) and (\ref{CUNTZ1}), $\RAN\Psi_\3(W)$ is also \iRR for
$\pi_{P_1}\0\R{W}$. Since both spaces contain $\phi_\3^W$, they must
coincide. By Lemma~\ref{lem:DEC}, $\bigoplus_\3\RAN\Psi_\3(W)=\FA{\KK_1}$,
i.e.~(\ref{CUNTZ2}) holds for $W$. 

Now let $\Psi_0(U)$ be the \UNI y \IMP er for $\8_U$ given by
(\ref{PNV}). Since $\R{V}=\8_U\R{W}$, the \ISM ies
$(\Psi_0(U)\Psi_\3(W))_{\3\in I_m}$ \iMP $\R{V}$ in $\pi_{P_1}$. We
are going to show that actually 
\begin{equation}
  \label{PVUV}
  \Psi_0(U)\Psi_\3(W)=\Psi_\3(V)
\end{equation}
holds under the above assumptions. Since each \IMP er is completely
determined by its value on $\Omega_{P_1}$ (this follows from
(\ref{CUNTZ1})), it suffices to prove (\ref{PVUV}) on
$\Omega_{P_1}$. Note that $[\Psi_0(U),\PME]=0$ since $\ker
U_{11}=\{0\}$. Hence $\Psi_0(U)\psi(f_j)=\psi(k_j)\Psi_0(U)$ by
$Uf_j=k_j$, and $[\Psi_0(U),A_{V,r}]=0$ by
$[\Psi_0(U),a_V(e_r)]=0$. Moreover, $\LU_{12}=\LV_{12}$ implies 
$$\Psi_0(U)\Omega_{P_1}=\PV\Omega_{P_1}$$
(see~(\ref{PV})), and we obtain 
\begin{equation} \label{IMPVAC}
  \Psi_0(U)\Psi_\3(W)\Omega_{P_1}=(-1)^{L_V}\psi(k_{\3_1})
  \cdots\psi(k_{\3_r})A_{V,1}\cdots
  A_{V,L_V}\PV\Omega_{P_1}=\Psi_\3(V)\Omega_{P_1}. 
\end{equation}
\hspace*{\fill} $\Box$
\section{Structure of the Semigroup of Implementable Endomorphisms}
\label{sec:STRUC}
Let $P_1$ be a \bP of $(\KK,\7)$ and $P_2:=\4{P_1}$. It is easily seen
that $\IK{P_1}$ is a topological \sG relative to the metric
(cf.~\cite{A87})
$$\delta_{P_1}(V,V'):=\|V-V'\|+\|V_{12}-V'_{12}\|_2.$$ The present
section is devoted to the study of the \CON ed components of
$\IK{P_1}=\bigcup_m\INK{2m}{P_1}$. It is inspired by the work of
Carey, Hurst and O'Brien~\cite{CHOB}.

Araki~\cite{A87} has shown that the group $\INK{0}{P_1}\subset
\IK{P_1}$ consists of two \CON ed components $\INK{0}{P_1}^\pm$.
Namely,
$$\chi(U):=(-1)^{\dim\ker U_{11}}$$ defines a continuous character
$\chi$ on $\INK{0}{P_1}$, and $\chi|_{\INK{0}{P_1}^\pm}=\pm1$.
However, we shall see that $\INK{2m}{P_1}$ is \CON ed if $m\neq0$ and
that $\chi:V\mapsto(-1)^{\dim\ker V_{11}}$ remains neither
multiplicative nor continuous when extended to the whole \sG
$\IK{P_1}$.

We need a preparatory result. Let $\HH$ be an infinite--\DIM al complex
\HSP. We prove that the subsets of $\B{\HH}$, consisting of \ISM ies
with fixed index, are \CON ed.
\begin{Lem}
  \label{lem:CONN}
  The sets ${\cal I}^n(\HH):=\{V\in\B{\HH}\ |\ V^*V=\1,\ \IND V^*=n\}$
  are arcwise \CON ed in the norm topology.
\end{Lem}
\begin{Proof}
  Let $V,V'\in {\cal I}^n(\HH)$. Since $\dim\ker V^*=\dim\ker V'^*$,
  there exists a \UNI y \oP $U$ on $\HH$ with $V'=UV$ (choose a
  partial \ISM y $u$ with initial space $\ker V^*$ and final space
  $\ker V'^*$ and set $U:=V'V^*+u$). Since the \UNI y group $\cal
  U(\HH)$ is arcwise \CON ed, there exists a \cOC $U(t)$ in $\cal
  U(\HH)$ with $U(0)=\1$ and $U(1)=U$. Then $U'(t):=U(t)V$ is a \cOC
  in ${\cal I}^n(\HH)$ with $U'(0)=V$ and $U'(1)=V'$.
\end{Proof}

Let us return to $\IK{P_1}$. In the following, the shorthand $V\sim
V'$ stands for the existence of a \cOC in $\IK{P_1}$ which \CON s $V$
to $V'$. Note that ``$\sim$'' is an equivalence relation and that
$V\sim V'$ implies $VV''\sim V'V''$ and $V''V\sim V''V'$ for
$V,V',V''\in\IK{P_1}$. 
\begin{Th}\label{th:CONN}
  The \CON ed components of $\IK{P_1}$ are precisely the sets
  $\INK{0}{P_1}^\pm$ and $\INK{2m}{P_1},\ 1\leq m\leq\infty$.
\end{Th}
\begin{Proofspc}
  Let $V\in\INK{2m}{P_1}$, and let $V=UW$ be a \dEC as in
  Proposition~\ref{prop:VUW}. Then $U\in\INK{0}{P_1}^+$ since $\ker
  U_{11}=\{0\}$. This implies $U\sim\1$ by Araki's result, hence
  $V=UW\sim W$.

  Since $V_{11}$ and ${V_{11}}^*$ both have infinite rank and
  since $\dim\ker V_{11}=\dim(V_{12}(\ker V_{22}))$
  (cf.~(\ref{LVFIN}), (\ref{KER})), there exists an \ISM y $\hat
  W_{11}$ on $\KK_1$ with
  $\IND\hat W_{11}{}^*=\dim(\ker{V_{11}}^*\ominus V_{12}(\ker
  V_{22}))=m$ (cf.~(\ref{DIMKER})) and 
  $$ \hat W_{11}(\RAN{V_{11}}^*)=\RAN V_{11},\qquad
     \hat W_{11}(\ker{V_{11}})=V_{12}(\ker V_{22}).$$
  Let $\hat W:=\hat W_{11}+\7\hat W_{11}\7\in\INK{2m}{P_1}$ be the
  \ASS ed \bOG\oP with $\hat W\hat W^*=WW^*$. Inserting the
  definitions, we find that 
  $$\hat U:=\hat W^*W\in\INK{0}{P_1}$$ is a \UNI y \bOG\oP with $\hat
  W\hat U=W$ and $\ker \hat U_{11}=\ker V_{11}$, hence $\chi(\hat
  U)=\chi(V)$.

  Now let $V'\in\INK{2m}{P_1}$ be another \bOG\oP with \cOR\OP s
  $W',\hat W'\in\INK{2m}{P_1},\ \hat U'\in\INK{0}{P_1}$. By
  Lemma~\ref{lem:CONN}, $\hat W\sim\hat W'$ since both are diagonal.
  Assume that $\chi(V)=\chi(V')$. Then $\hat U\sim\hat U'$ by Araki's
  result, and we conclude 
  $$V\sim W=\hat W\hat U\sim\hat W\hat U'\sim\hat W'\hat U'=W'\sim
  V'.$$ Therefore either of the two subsets
  $\INK{2m}{P_1}^\pm:=\{V\in\INK{2m}{P_1}\ |\ \chi(V)=\pm1\}$ is
  arcwise \CON ed. Below, we give an example of a \cOC in
  $\INK{2m}{P_1}$ which \CON s $\INK{2m}{P_1}^+$ to
  $\INK{2m}{P_1}^-$. Hence $\INK{2m}{P_1}$ itself is \CON ed.
  Of course, $V\sim V'$ cannot hold if $\IND V\neq\IND V'$.
\end{Proofspc}
{\it Example}. Let $V(\PHI)$ be the \bOG\oP introduced in the example
in Section~\ref{sec:DECREP} (with $P=P_1$). Then
$V(\PHI)\in\INK{2}{P_1}$ since
${V(\PHI)_{12}}^*V(\PHI)_{12}=(S_{V(\PHI)})_{22}=(1-\9_\PHI)\4{E_0}$
has finite rank, and $\PHI\mapsto V(\PHI)$ is a \cOC in
$\INK{2}{P_1}$. We have $\ker V(\PHI)_{11}=\ker
(S_{V(\PHI)})_{11}=\ker(\9_\PHI E_0+\sum_{n\geq1}E_n)$, hence
$$\chi(V(\PHI))=\left\{\begin{array}{rl}
  1, & \PHI\notin(4\ZZ+3)\pi/4 \\
  -1, & \PHI\in(4\ZZ+3)\pi/4. 
\end{array}\right.$$
Let $V\in\INK{2m-2}{P_1}$ with $[P_1,V]=0$. Then
$\chi(VV(\PHI))=\chi(V(\PHI))$ since $V_{11}$ is \ISM ic, so
$\PHI\mapsto VV(\PHI)$ is a \cOC in $\INK{2m}{P_1}$ which \CON s
$\INK{2m}{P_1}^+$ to $\INK{2m}{P_1}^-$. This completes the proof of
Theorem~\ref{th:CONN}.

$V(\PHI)$ may also serve to illustrate that $\chi$ is not
multiplicative on $\IK{P_1}$. 
Define a \bOG\oP
$U:=2^{-1/2}f_0^+\<{f_0^++f_1^-,.\,}-2^{-1/2} f_1^+
\<{f_0^--f_1^+,.\,}+2^{-1/2}f_0^-\<{f_0^-+f_1^+,.\,}
-2^{-1/2}f_1^-\<{f_0^+-f_1^-,.\,} +\sum_{n\geq2}(E_n+\4{E_n})$.
Then $U\in\INK{0}{P_1}$, and a calculation shows that
$U_{11}=\frac{1}{\sqrt{2}}(E_0+E_1)+\sum_{n\geq2}E_n$ and
$UV(\frac{3\pi}{4})=V(\frac{\pi}{2})$. This entails
$$1=\chi\Big(UV(\mbox{$\frac{3\pi}{4}$})\Big)\neq
\chi(U)\chi\Big(V(\mbox{$\frac{3\pi}{4}$})\Big)=-1$$ since $\ker
U_{11}=\ker V(\frac{\pi}{2})_{11}=\{0\}$, but $\ker
V(\frac{3\pi}{4})_{11}=\CC f_0$. 
We finally note that the eigenvalues
$\pm(1-\9_\PHI)$ of $P_1-S_{V(\PHI)}=(1-\9_\PHI)(E_0-\4{E_0})$ have
multiplicity one if $\9_\PHI\neq1$, in contrast to the \UNI y case
where the multiplicities of eigenvalues in (0,1) are always
even~\cite{A87}.
\\[2ex]{\em Acknowledgments. } I would like to thank Professor
K.~Fredenhagen for the initiation of this work and for many helpful
discussions. Thanks for discussions are also due to J.~B\"ockenhauer
and to my colleagues at the FU Berlin.

\end{document}